\documentclass[10pt,journal,compsoc]{IEEEtran}

\usepackage{subfigure}
\usepackage{color}
\usepackage{colortbl}
\usepackage{pifont}
\usepackage{tikz}
\newcommand*{\circled}[1]{\lower.7ex\hbox{\tikz\draw (0pt, 0pt)%
    circle (.5em) node {\makebox[1em][c]{\small #1}};}}
    
\newcommand{\toolname}{\textsc{\textbf{Scan}}}

\usepackage{ragged2e}

\usepackage[normalem]{ulem}

\newcommand{\revise}[1]{{\color{black}{#1}}}
\newcommand{\delete}[1]{}

\newcommand{\newrevise}[1]{{\color{black}{#1}}}
\newcommand{\newdelete}[1]{}

\usepackage{url}
\usepackage{hyperref}
\hypersetup{
    colorlinks=true,
    linkcolor=blue,
    filecolor=blue,      
    urlcolor=blue,
    citecolor=blue,
}

\usepackage{booktabs}

\usepackage{algorithmic}
\usepackage{algorithm}

\usepackage{multirow}

\usepackage{bm}%
\usepackage{enumitem} %

\usepackage{amsmath,amsthm,amssymb,amsfonts}
\newtheorem{definition}{Definition}
\usepackage{bbm}

\usepackage{blindtext}
\newenvironment{mylist}[1]%
{\begin{list}{}{\settowidth{\labelwidth}{\bf #1}%
			\setlength{\leftmargin}{\labelwidth}%
			\addtolength{\leftmargin}{\labelsep}%
			}}%
{\end{list}}

\usepackage[most]{tcolorbox}
\newcommand{\myfinding}[2]{
\begin{center}
\begin{tcolorbox}[colback=gray!15, colframe=black, boxsep= -0.15cm, middle=-0.15cm, breakable]
\textbf{Answer to RQ{#1}:}
{#2}
\end{tcolorbox}
\end{center}
}

\newcommand{\mydiscussion}[1]{
\begin{center}
\begin{tcolorbox}[colback=gray!15, colframe=black, boxsep= -0.15cm, middle=-0.15cm, breakable]
\textbf{Summary:}
{#1}
\end{tcolorbox}
\end{center}
}

\ifCLASSOPTIONcompsoc
  \usepackage[nocompress]{cite}
\else
  \usepackage{cite}
\fi

\begin{document}

\title{Pre-trained Model-based Automated Software Vulnerability Repair: How Far are We?}

\author{Quanjun Zhang, Chunrong Fang, Bowen Yu, Weisong Sun, Tongke Zhang, Zhenyu Chen
        
\IEEEcompsocitemizethanks{
\IEEEcompsocthanksitem 
Quanjun Zhang, Chunrong Fang, Bowen Yu, Weisong Sun, Tongke Zhang and Zhenyu Chen
are with the State Key Laboratory for Novel Software Technology, Nanjing University, China. \protect\\
E-mail: 
\{quanjun.zhang, 201250070, weisongsun, 201250032\} @smail.nju.edu.cn,
\{fangchunrong, zychen\}@nju.edu.cn.

\IEEEcompsocthanksitem Chunrong Fang and Zhenyu Chen are the corresponding authors.
}
\thanks{Manuscript received xxx xxx, 2023; revised xxx xxx, 2023.}
}

\markboth{Journal of \LaTeX\ Class Files,~Vol.~14, No.~8, August~2023}%
{Shell \MakeLowercase{\textit{et al.}}: Bare Demo of IEEEtran.cls for Computer Society Journals}

\IEEEtitleabstractindextext{%
\begin{abstract}
\justifying
Various approaches are proposed to help under-resourced security researchers to detect and analyze software vulnerabilities.
It is still incredibly time-consuming and labor-intensive for security researchers to fix such reported vulnerabilities due to the increasing size and complexity of modern software systems.
The time lag between the reporting and fixing of a \revise{security} vulnerability causes software systems to suffer from significant exposure to possible attacks.
Very recently, some techniques propose to apply pre-trained models to fix security vulnerabilities and have proved their success in improving repair accuracy.
However, the effectiveness of existing pre-trained models has not been systematically compared and little is known about their advantages and disadvantages.

To bridge this gap, we perform the first extensive study on applying various pre-trained models to automated vulnerability repair.
The experimental results on two vulnerability datasets show that all studied pre-trained models consistently outperform the state-of-the-art technique VRepair with a prediction accuracy of 32.94\%$\sim$44.96\%.
We also investigate the impact of three major phases (i.e., data pre-processing, model training and repair inference) in the vulnerability repair workflow.
Inspired by the findings, we construct a simplistic vulnerability repair approach that adopts the transfer learning from bug fixing. 
Surprisingly, such a simplistic approach can further improve the prediction accuracy of pre-trained models by \delete{10.90}\revise{9.40}\% on average.
Besides, we provide additional discussion from different aspects (e.g., \delete{the repaired vulnerability CWE types and the used code representation}\revise{code representation and a preliminary study with ChatGPT}) to illustrate the capacity and limitation of pre-trained model-based techniques.
Finally, we further pinpoint various practical guidelines (e.g., \delete{the usage of pre-trained models}\revise{the improvement of fine-tuning}) for advanced pre-trained model-based vulnerability repair in the near future.
Our study highlights the promising future of adopting pre-trained models 
to patch real-world security vulnerabilities and reduce the manual debugging effort of security experts in practice.

\end{abstract}

}

\maketitle

\IEEEdisplaynontitleabstractindextext

\IEEEpeerreviewmaketitle

\section{Introduction}
\IEEEPARstart{S}{}oftware vulnerabilities generally refer to security flaws in source code and are prevalent in modern software system evolution \cite{shahzad2019large}.
Attackers can utilize unresolved vulnerabilities to undertake malicious activities, posing enormous risks to millions of users around the globe \cite{dowd2006art}.
For example, the \textit{Log4Shell} vulnerability (CVE-2021-44228) from Apache Log4j library \cite{log4j} allows attackers to run arbitrary code on any affected system \cite{Log4Shell} and is widely recognized as the most severe vulnerability in the last decade (e.g., 93\% of the cloud enterprise environment are vulnerable to \textit{Log4Shell} \cite{Log4Shell2}).
Nowadays, the number of exposed security vulnerabilities recorded by the National Vulnerability Database
(NVD) \cite{NVD} has been increasing at a striking speed \delete{(e.g., already exceeding forty thousand in the last two years, shown in Fig.~\ref{fig:cve})}, affecting millions of software systems annually.

Various approaches have been proposed to help security experts to understand the characteristics of vulnerabilities better (e.g., vulnerability analysis \cite{yan2018spatio}) and find vulnerabilities faster (e.g., vulnerability detection \cite{li2021vuldeelocator, bohme2016coverage, chakraborty2021deep}) from both academia and industry.
For example, static analysis tools (e.g., Infer \cite{infer}) aim to identify suspicious code lines containing a vulnerability based on pattern matching or data/control flow analysis \cite{pornprasit2022deeplinedp},
while fuzzing tools (e.g., AFL \cite{zhu2022fuzzing}) expose vulnerabilities by dynamically executing inputs that trigger potentially exploitable issues.
Such vulnerability hunters only help security experts to detect and localize the location of security vulnerabilities.
However, it is incredibly time-consuming and labor-intensive for security experts to repair such security vulnerabilities manually due to the strikingly increasing number of detected vulnerabilities and the complexity of modern software systems \cite{zhang2022program,gao2021beyond}.
For example, previous studies report that the average time for repairing severe vulnerabilities is 256 days \cite{fixcve} and the life spans of 50\% of vulnerabilities even exceed 438 days \cite{li2017large}.
It is extremely time-critical to patch reported security vulnerabilities as a belated vulnerability repair could expose software systems to attack \cite{liu2020large,li2021sysevr}, damaging hundreds of millions of dollars to the global economy yearly.

Recently, Chen et al. \cite{chen2022neural} propose VRepair, an automated software vulnerability repair approach that leverages a transformer neural network model.
VRepair employs transfer learning to address the problem that existing vulnerability-fixing datasets are too small to be meaningfully used in a deep learning model.
In VRepair, transfer learning means acquiring generic knowledge from a bug fixing dataset and then transferring the learned knowledge to repair security vulnerabilities.
Despite its advanced design, the effectiveness of VRepair is still limited because the transfer learning is built upon a small bug fixing corpus with only 23,607 C/C++ functions.
Thus, the actual advantage offered by transfer learning on the critical automated vulnerability repair problem is yet to be revealed.

\textbf{This paper.}
Recent years have seen the successful application of transfer learning in large pre-trained models.
Different from VRepair, such models are first pre-trained by self-supervised training on a large-scale unlabeled corpus (e.g., CodeSearchNet with 2.3 million functions) to derive generic language representation, and then transferred to benefit multiple downstream tasks by fine-tuning on a limited labeled corpus.
Pre-trained models have greatly boosted the performance in a broad spectrum of code-related tasks (e.g., code summarization) \cite{wang2021codet5,feng2020codebert}.
However, the literature does not yet provide extensive experimental results to understand the actual benefits of pre-trained models when dealing with the automated software vulnerability repair task.
In this paper, to fill this gap,  we investigate the feasibility of leveraging advances in pre-trained models to repair security vulnerabilities and facilitate future vulnerability repair studies.
We address the following five research questions:
RQ1 answers the repair performance; 
RQ2$\sim$RQ4 analyze different aspects of the repair workflow and
RQ5 explores a conceptually simple repair technique.

\begin{mylist}{\textbf{(RQ1)}}
  \item[\textbf{(RQ1)}] \textbf{The effectiveness of state-of-the-art pre-trained model-based vulnerability repair techniques.}
  
  \textbf{\underline{Results:} }
  pre-trained models achieve 32.93\%$\sim$44.96\% prediction accuracy, which is 10.21\%$\sim$22.23\% more than VRepair, indicating the substantial benefits of using \delete{pre-training~}\revise{pre-trained} models for vulnerability repair.
  
  \item[\textbf{(RQ2)}] \textbf{The influence of code context, abstraction and tokenization methods in the pre-processing phase.}
  
  \textbf{\underline{Results:}}
 (1) code context improves the prediction accuracy by 4.51\% for CodeT5, while other models have a 7.68\%$\sim$15.71\% accuracy increase without code context;
 (2) subword tokenization improves the prediction accuracy by \delete{12.07}\revise{15.69}\% on average compared with word-level tokenization, highlighting the substantial benefits of using subword tokenization for vulnerability repair approaches;
 (3) raw source code improves the prediction accuracy by \delete{18.22}\revise{18.26}\% on average compared with abstracted source code, demonstrating the substantial benefits of meaningful vector representation in pre-trained models.
  
  \item[\textbf{(RQ3)}] 
  \textbf{The role of the pre-training component and fine-tuning component in the model training phase.}
  
  \textbf{\underline{Results:}}
  (1) without a pre-training component, the prediction accuracy decreases by 12.02\%$\sim$\delete{22.51}\revise{33.36}\%, highlighting the substantial benefits of the pre-training on a large corpus;
  (2) the size of the fine-tuning training dataset plays a crucial role in vulnerability repair, e.g., each 20\% of the training dataset brings about a 10\% increase in prediction accuracy.
  
  \item[\textbf{(RQ4)}] 
  \textbf{The influence of beam size and \delete{vulnerability size}\revise{function length} variables in the repair inference phase.}
  
  \textbf{\underline{Results:} }
  (1) when considering smaller beam size, pre-trained models achieve comparable or better prediction accuracy than VReapir, such as 42.38\% with Top-5 candidates for CodeT5, highlighting the practicality in human inspection scenario;  
  (2) when considering larger beam size, the prediction accuracy still increases, such as 45.31\% with Top-200 candidates for CodeT5, highlighting the superiority in automated inspection scenario;
  (3) the \delete{vulnerability size}\revise{function length} has a crucial influence on the prediction accuracy due to the inherent limitation of the transformer, e.g., prediction accuracy \revise{of CodeT5} decreases from 56.80\% to 30.90\% if the \delete{vulnerability size}\revise{vulnerable function} exceeds 500 \revise{code} tokens\delete{.}\revise{;}
  \revise{(4) pre-trained models can accurately repair many real-world and dangerous vulnerabilities (e.g., NULL Pointer Dereference and Improper Input Validation).}

  \item[\textbf{(RQ5)}]
  \textbf{Enhanced pre-trained model-based vulnerability repair technique via bug fixing transfer learning.}
   
  \textbf{\underline{Results:}}
  transfer learning from bug fixing further boosts the repair performance of pre-trained models, such as a new high prediction accuracy of 53.75\% by UniXcoder, indicating the potential ability of generic knowledge from bug fixing for vulnerability repair.
  
\end{mylist}

We also provide additional discussion to demonstrate that 
\delete{(1) pre-trained models can accurately repair many real-world and dangerous vulnerabilities (e.g., NULL Pointer Dereference and Improper Input Validation);}
\delete{(2)}\revise{(1)} improper usage of some tags may result in adverse effects for pre-trained models;
\delete{(3)}\revise{(2)} prompt code representation can further improve the repair performance of pre-trained models\delete{.}\revise{;}
\revise{(3) ChatGPT fails to provide remarkable repair performance in a zero-shot scenario despite its extensive programming language knowledge.}
Finally, based on our findings, we pinpoint open research challenges and provide practical guidelines for applying pre-trained models to vulnerability repair for future studies.
Overall, our results lead us to conclude that pre-trained models are considerably accurate and practical repair approaches, highlighting the substantial advancement of pre-trained models on automated vulnerability repair.

\textbf{Novelty \& Contributions.}
To sum up, the main contributions of this paper are as follows:
\begin{itemize}
\item
\textit{\textbf{New Dimension\revise{.}}}
We bridge the gap between the recent advances in pre-trained models and a crucial software security problem, i.e., \textit{automated vulnerability repair}. 
Our work not only provides an extensive evaluation of the recent pre-trained models, but also reveals the potential for leveraging pre-trained models to resolve the important automated vulnerability repair problem.

\item
\textit{\textbf{Extensive Study.}}
We conduct the first large-scale empirical study of how pre-trained models are applied to automated software vulnerability repair, involving more than 100 variants of models.
Specifically, our study contains 
(1) a systematic comparison between studied pre-trained models and the state-of-the-art technique VRepair, indicating that such simple pre-trained model-based techniques can easily outperform VRepair;
(2) an extensive evaluation to analyze the impact of different aspects in the typical vulnerability repair workflow;
(3) an additional discussion about the CWE type of repaired vulnerability and the code representation, etc.

\item
\textit{\textbf{Effective Technique.}}
We propose a conceptually simple pre-trained model-based vulnerability repair approach, which simply adopts the transfer learning paradigm from bug fixing to boost vulnerability repair further.

\item
\textit{\textbf{Practical Guidelines.}}
We provide several practical guidelines on applying pre-trained models for future software vulnerability repair studies, such as exploring code representation and utilizing bug fixing capability.

\end{itemize}  

\textbf{Open Science.}
To support the open science community, we release the studied dataset, scripts (i.e., data processing, model training, and model evaluation), and related models in our experiment for replication and future research \cite{myurl}.

\delete{
\textbf{Paper Organization.} 
The rest of this paper is organized as follows.
Section \ref{sec:bk} reviews some background information.
Section \ref{sec:study} introduces our empirical study design.
Section \ref{sec:re&an} provides the detailed results of our empirical study and answers the research questions.
Section \ref{sec:dis} presents an additional discussion.
Section \ref{sec:implication} provides practical guidelines.
Section \ref{sec:threats} discloses the threats to validity and Section \ref{sec:related} presents related work.
Section \ref{sec:conclusion} draws the conclusions.
}

\section{Background}
\label{sec:bk}
\subsection{Software Vulnerability}
Nowadays, thousands of vulnerabilities are reported to vulnerability databases monthly, such as the Common Vulnerability Exposure (CVE) \cite{cve} and National Vulnerability Database (NVD) \cite{NVD}.
In the vulnerability databases, each vulnerability is assigned a unique identifier (i.e., CVE ID).
Each vulnerability with a CVE ID is also assigned to a Common Weakness Enumeration (CWE) category representing the generic type of the weakness, defined as follows:

\begin{definition}\textbf{CVE:} 
a CVE ID identifies a vulnerability within
the Common Vulnerabilities and Exposures database. 
It is a unique alphanumeric assigned to a specific security vulnerability.
\end{definition}

For example, the entry identified as CVE-2021-44228 is a Remote Code Execution (RCE) vulnerability in Apache Log4j due to a Java Naming and Directory Interface (JNDI) injection.
The term 2021 is the year in which the CVE ID is assigned or the vulnerability is made public.
The term 44228 is a unique identifier of the vulnerability within the year.

\begin{definition}{\textbf{CWE:}}
A CWE ID is a Common Weakness Enumeration and identifies the category that a CVE ID is a part of.
CWE is a community-developed list of software and hardware weakness types and represents general software weaknesses.
\end{definition}

For example, CVE-2021-44228 is categorized into CWE-917, representing the Expression Language Injection category.
Among all CWE types, CWE-787 (Out-of-bounds Write), CWE-79 (Cross-site Scripting) and CWE-89 (SQL Injection) are the Top-3 most dangerous weaknesses \cite{top25}.

Given the potentially disastrous effect when software vulnerabilities are exploited, a mass of work in software engineering and security communities is conducted on early vulnerability detection \cite{tantithamthavorn2016automated, movahedi2020predicting} and more recently vulnerability repair \cite{fu2022vulrepair,chen2022neural}, which is the focus of this paper.

\subsection{Automated Vulnerability Repair}
Automated vulnerability repair aims to fix security vulnerabilities automatically without human debugging effort and plays a crucial role in countering emergent security concerns \cite{noller2022trust, belleville2019kald} in the security community.
The typical workflow of vulnerability repair is usually composed of four steps:
(1) \textit{the detection phase} utilizes vulnerability analysis tools under consideration (e.g., Infer \cite{infer} and SpotBugs \cite{spotbugs}) to statically  analyze source code and indicate a suspicious line containing a vulnerability;
(2) \textit{the repair phase} then generates various new program variants (i.e., candidate patches) by applying a set of transformation rules to these suspicious lines \cite{chen2022neural};
(3) \textit{the verification phase} recompiles the vulnerable program with the generated patch to check if it can pass a functional test suite (if such a test suite exists), and then adopts the vulnerability analysis tools (e.g., Infer \cite{infer}) to check if the reported vulnerability is fixed;
(4) \textit{the deployment phase} employs security researchers to review and deploy a patch in practice after the patch has been found to pass the test suite and vulnerability analysis tools \cite{duan2019automating,zhang2022program1}.

Existing studies have demonstrated that manually repairing security vulnerabilities is extraordinarily time-consuming and a delayed repair may expose software systems to attacks.
Inspired by the success of transformers with an encoder stack and decoder stack in code generation tasks (e.g., code summarization), vulnerability repair can be formulated as a neural machine translation task.
NMT-based vulnerability repair problem is formally defined as follows:

\begin{definition}\textbf{NMT-based Vulnerability Repair:} 
Given a vulnerable function $X_i=\left[x_1, \ldots, x_n\right]$ with $n$ code tokens and vulnerability repair $Y_i=\left[y_1, \ldots, y_m\right]$ with $m$ code tokens, the problem of vulnerability repair is to maximize the conditional probability: $P\left(Y \mid X \right) = \prod_{i=1}^{m} P\left(y_i \mid y_1, \ldots, y_{i-1}; x_1, \ldots, x_n\right)$.
\end{definition}

In other words, the objective of an NMT-based vulnerability repair model is to learn the mapping between a vulnerable function $X$ and a vulnerability repair $Y$.
Then the parameters of the model are updated by using the training dataset, so as to optimize the mapping (i.e., maximizing $P$).

\subsection{Pre-trained Model}
Pre-trained models have significantly improved performance across a wide range of natural language processing (NLP) and code-related tasks\, such as machine translation, defect detection, and code classification~\cite{lu2021codexglue,guo2020graphcodebert}.
Typically, the models are pre-trained to derive generic vector representation by self-supervised training on a large-scale unlabeled corpus and then are transferred to benefit multiple downstream tasks by fine-tuning on a limited labeled corpus \cite{devlin2018bert}.

\begin{figure*}[htbp]
    \centering
    \includegraphics[width=0.9\linewidth]{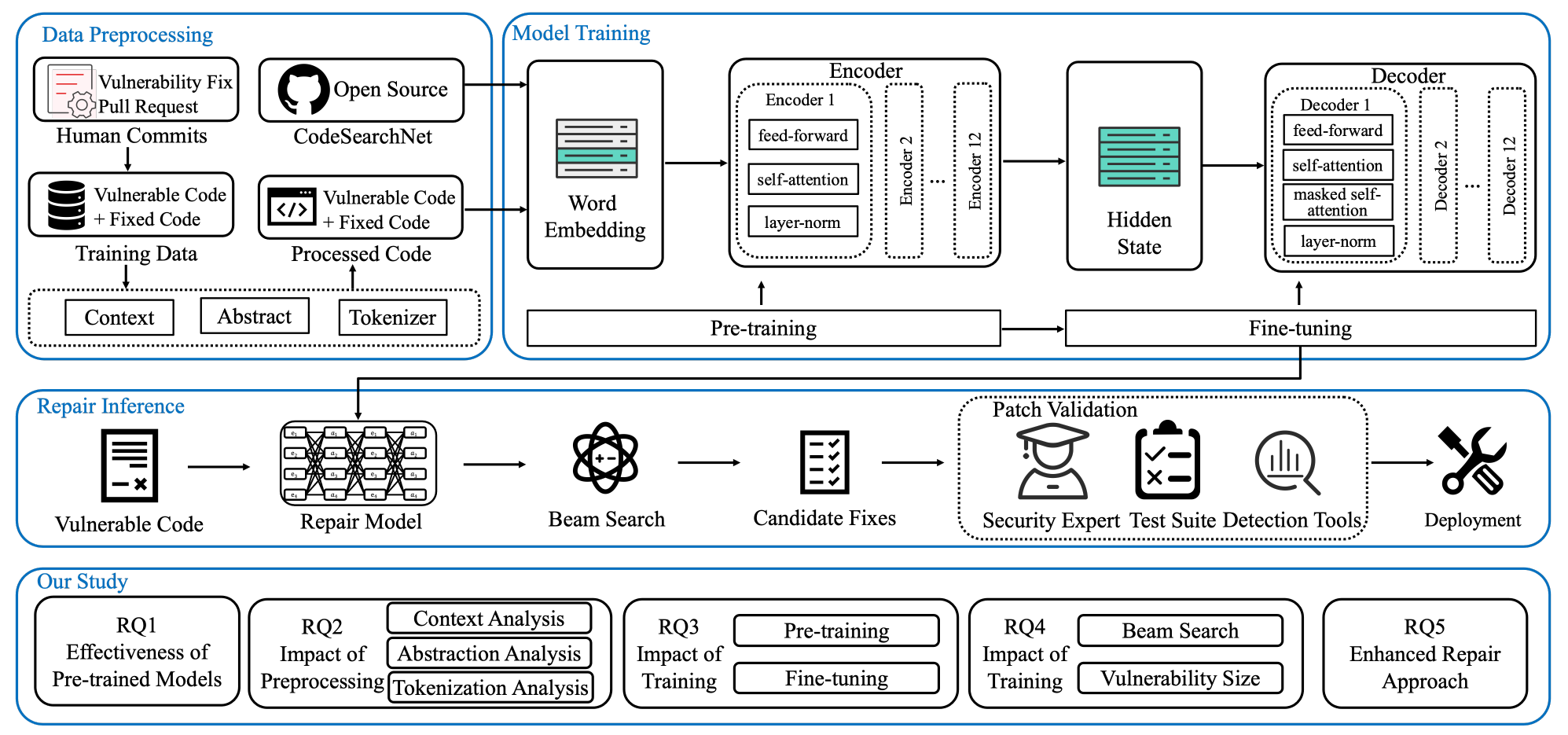}
    \caption{\revise{Overview of our empirical study.}}
    \label{fig:workfolw}
\end{figure*}

Existing pre-trained models generally adopt the encoder-decoder transformer architecture, which can be classified into three types: encoder-only, decoder-only, and encoder-decoder models.
Encoder-only models (e.g., CodeBERT \cite{feng2020codebert}) usually pre-train a bidirectional transformer where tokens can attend to each other.
Encoder-only models are good at understanding tasks (e.g., code search), but their bidirectionality nature requires an additional decoder for generation tasks.
Decoder-only models (e.g., GPT \cite{brown2020gpt}) are pre-trained using unidirectional language modeling that only allows tokens to attend to the previous tokens and themselves to predict the next token.
Decoder-only models are good at auto-regressive tasks like code completion, but the unidirectional framework is sub-optimal for understanding tasks.
Encoder-decoder models (e.g., T5 \cite{raffel2019t5}) often make use of denoising pre-training objectives that corrupt the source input and require the decoder to recover them.
Compared to encoder-only and decoder-only models that favor understanding and auto-regressive tasks, encoder-decoder models can support generation tasks well like code summarization.

In the context of vulnerability repair, an encoder stack takes a sequence of code tokens as input to map a vulnerable function $X_i=\left[x_1, \ldots, x_n\right]$ into a fixed-length intermediate hidden state, while the decoder stack takes the hidden state vector as an input to generate the output sequence of tokens $Y_i=\left[y_1, \ldots, y_n\right]$.
Despite promising results in previous code-related work, little work attempts to explore the capabilities of pre-trained models in supporting automated vulnerability repair.
In this work, \delete{four}\revise{five} pre-trained models are exploited (discussed in Section \ref{sec:model}) to repair vulnerabilities, as they have been widely adopted in various code-related tasks and are pretty effective for code generation tasks \cite{guo2020graphcodebert}.

\section{Study on Software Vulnerability Repair}
\label{sec:study}

\subsection{Overview}

Fig. \ref{fig:workfolw} shows a general workflow of the existing NMT-based automated vulnerability repair framework when pre-trained models are integrated.
The typical workflow can be divided into three steps\delete{ (i.e., data pre-processing, model training and repair inference)}, illustrated as follows.

\textbf{\ding{172} \text{In the data pre-processing phase}}, a given software vulnerability code snippet (e.g., vulnerable function) is taken as the input and the processed code tokens are returned.
According to existing NMT-based repair studies \cite{chen2022neural,chi2022seqtrans},  there generally exist three potential ways to \delete{prepossess~}\revise{preprocesses} the vulnerable function: code context, abstraction and tokenization.
First, \textit{code context} information refers to other correlated non-vulnerable lines within the vulnerable function.
Previous work has demonstrated that NMT-based repair models reveal diverse code changes to fix bugs under different contexts \cite{chen2019sequencer}.
Second, \textit{code abstraction} renames some special words (e.g., string and number literals) to a pool of predefined tokens.
Code abstraction has been proven to be a significant method in reducing the size of the vocabulary.
Third, \textit{code tokenization} splits source code into words or subwords, which are then converted to ids through a look-up table.
Pre-trained models generally perform subword tokenization to split the vulnerable function into meaningful subwords while preserving the common tokens well.

\textbf{\ding{173} \text{In the model training phase}}, an NMT-based repair model is first built on the top of the transformer \cite{vaswani2017attention} and the mapping from vulnerable code to fixed code is optimized by updating the parameters of the designed model.
Similar to the vanilla transformer architecture \cite{vaswani2017attention}, pre-trained models usually start with an encoder stack and a decoder stack, and end with a linear layer with softmax activation.
The pre-trained models perform a word embedding to generate representation vectors for tokenized vulnerable functions.
Then pre-trained models feed the vectors to an encoder stack to derive the \delete{encoder }hidden state, which are further passed into a decoder stack.
The output of the decoder stack is finally fed into a linear layer with softmax activation to generate the probability distribution of the vocabulary.

\textbf{\ding{174} \text{In the repair inference phase}}, after the NMT-based repair model is well trained, the beam search strategy is leveraged to generate the candidate patches as prediction results based on the probability distribution of the vocabulary.
The generated candidate patches are then verified by the available functional test suite, vulnerability analysis tools or security experts for deployment in the software pipeline.

\subsection{Research Questions}

\delete{
We design our empirical study by asking research questions revolving around different aspects of pre-trained model-based automated software vulnerability repair.}
\revise{In this work, we explore the following research questions.}

\begin{description}
	
  \item [RQ1:] What is the performance of pre-trained models on automated software vulnerability repair when compared with the state-of-the-art technique?

  \item [RQ2:] How do the dataset pre-processing methods affect the performance of pre-trained models on automated software vulnerability repair?
        \begin{description}
        \item [RQ2.1:] What is the impact of \delete{the }code context?
        \item [RQ2.2:] What is the impact of code abstraction?
        \item [RQ2.3:] What is the impact of code tokenization?
    \end{description}
    
    \item [RQ3:] How do the model training components affect the performance of pre-trained models on automated software vulnerability repair?
    \begin{description}
        \item [RQ3.1:] What is the role of a pre-training component?
        \item [RQ3.2:] What is the role of a fine-tuning component?
    \end{description}
    
  \item [RQ4:] How do the model inference variables affect the performance of pre-trained models on automated software vulnerability repair?
    \begin{description}
        \item [RQ4.1:] What is the impact of beam size?
        \item [RQ4.2:] What is the impact of \delete{vulnerability size}\revise{function length}?
        \item [\revise{RQ4.3:}] \revise{What is the impact of CWE types?}
        
    \end{description}
    
  \item [RQ5:] Can we enhance the effectiveness of existing pre-trained model-based vulnerability repair techniques via transfer learning from bug fixing?

\end{description}

\subsection{Evaluation Metrics}
In our experiment, following existing vulnerability repair work \cite{chen2022neural,fu2022vulrepair}, we utilize exact match \textit{Accuracy@K} to evaluate the studied repair approaches.
\textit{Accuracy@K} measures the percentage of cases in which the vulnerability repair candidate predicted by the model equals the oracle repair sequence.
Since we use a beam-search decoding strategy, $k$ repair candidates with the highest probability are reported.

Given a vulnerability dataset $V = \{v_1, v_2, \cdots, v_n\}$ containing $n$ vulnerable functions, for a vulnerable function \delete{$v_k$~}\revise{$v_i$}, the beam-search decoding strategy generates a candidate set $C_i = \{c_{i}^1, c_{i}^2, \cdots, c_{i}^k \}$ containing $k$ potential vulnerability repair patches.
Then \textit{Accuracy@K} value is defined as follows:

\begin{equation}
	Accuracy@K = \frac
	{\sum_{i=1}^{n} \mathbbm{1} \{match(\sum_{j=1}^{k} c_i^j)\}}{n}
\end{equation}
where $\mathbbm{1}$ denotes whether $C_i$ contains a predicted repair sequence equal to the oracle repair sequence\revise{ and $K$ denotes the number of candidate patches}.
The sequence accuracy is 1 if any predicted sequence among the $k$ outputs matches the ground truth sequence, and it is 0 otherwise.

\subsection{Selected Models}
\label{sec:model}
\revise{
We select pre-trained models that satisfy the following criteria.
(1) the pre-trained language model is publicly accessible as we need to fine-tune the model, e.g., Codex~\cite{chen2021evaluating} and GPT-3~\cite{brown2020language} are excluded as their models and code are not released;
(2) the pre-trained model is trained on a large enough programming language corpus, e.g., we exclude T5~\cite{raffel2020exploring} and GPT-2~\cite{radford2019language}, which are natural language models.
As a result, we select five state-of-the-art pre-trained models (i.e., CodeBERT, GraphCodeBERT, CodeT5, UniXcoder, and CodeGPT), all of which are used in a recent study~\cite{guo2022unixcoder}.}

We describe the studied \delete{four}\revise{five} pre-trained models in our experiment.
The utilized models are from different architectures \revise{(e.g., encoder-decoder, encoder-only and decoder-only)} and organizations \revise{(e.g., Microsoft and Salesforce)} to obtain diverse evaluation scenarios.
\delete{
All models represent state-of-the-art and are publicly available on the Hugging Face, which is the largest open-source platform to host and deploy large models.}
\revise{All models represent state-of-the-art and are publicly available on the Hugging Face, the largest open-source platform to host large models.}
\delete{We summarize the selected pre-trained models as follows.}

\textbf{$\bullet$ CodeT5.}
Wang et al. \cite{wang2021codet5} present a pre-trained encoder-decoder model (\textit{CodeT5}) that considers the token type information in code based on T5 architecture.  
CodeT5 employs a unified framework to support both code understanding seamlessly and generation tasks\delete{ and allows for multi-task learning}.
\delete{
The most important feature of CodeT5 is that the code semantics of identifiers are taken into consideration in it. 
Assigned by developers, identifiers often convey rich code semantics and thus a novel identifier-aware objective is added to the training of CodeT5.}
\revise{
CodeT5 takes into account the code semantics of identifiers, adding a novel identifier-aware objective during its training, as these identifiers are assigned by developers and convey rich code semantics.}

\textbf{$\bullet$ CodeBERT.}
Feng et al. \cite{feng2020codebert} present a bimodal pre-trained model (\textit{CodeBERT}) for natural language and programming language with a transformer-based architecture.
CodeBERT utilizes two pre-training objectives (i.e., masked language modeling and replaced token detection) to support both code search and code documentation generation tasks.

\textbf{$\bullet$ GraphCodeBERT.}
Guo et al. \cite{guo2020graphcodebert} present the first pre-trained model (\textit{GraphCodeBERT}) that leverages code structure to improve code understanding tasks (e.g., code search and code translation).
Unlike existing models that take syntactic-level information (e.g., AST), GraphCodeBERT takes semantic-level information of code (e.g., data flow) for pre-training with a transformer-based architecture. 

\textbf{$\bullet$ UnixCoder.}
Guo et al. \cite{guo2022unixcoder} present a unified cross-modal pre-trained model (\textit{UniXcoder}) for programming languages to support both code-related understanding and generation tasks.
UniXcoder utilizes mask attention matrices with prefix adapters to control the behavior of the model and leverages cross-modal contents such as AST and code comment to enhance code representation.

\revise{\textbf{$\bullet$ CodeGPT.}
Lu et al.~\cite{lu2021codexglue} present a GPT-style Transformer-based pre-trained model (\textit{CodeGPT}) on top of the programming language.
CodeGPT has the same model architecture and training objective (i.e., predicting the next word given all of the previous words) as GPT-2, which consists of 12 layers of Transformer decoders.
}

\subsection{Dataset}
\label{sec:dataset}
In our experiment, following existing work \cite{chen2022neural,fu2022vulrepair}, we use two existing vulnerability datasets Big-Vul \cite{fan2020ac} and CVEFixes \cite{bhandari2021cvefixes} to evaluate the performance of studied vulnerability repair techniques.
The CVEfixes automatically fetches all available CVE records from the NVD, and gathers the vulnerable functions and corresponding fixes from associated open-source projects.
Similar to CVEfixes, the Big-Vul dataset crawls the public CVE database and collects the descriptive information of the vulnerabilities, e.g., CVE IDs, CVE summaries and CWE IDs.
With CVE information and its related published Github code repository links, Big-Vul identifies vulnerability-related code commits and extracts relevant code changes that fix the vulnerability.

To conduct a fair comparison, we directly reuse the replication package pre-processed by Chen et al. \cite{chen2022neural}.
The dataset is randomly divided into training, validation and testing sets, with 5936 (70\%), 839 (10\%) and 1706 (20\%) examples in each respective set.
We present the number of vulnerability samples for the Top-10 CWE values in Table \ref{tab:dataset}.
The software vulnerability is divided into an input sequence and its relevant output sequence.
The input sequence begins with a special tag that specifies the CWE type of the vulnerability, followed by multiple vulnerable code snippets.
Each vulnerable code snippet is identified with the special tags $\langle \texttt{StartLoc}\rangle$ and $\langle \texttt{EndLoc}\rangle$, where the former denotes the beginning of the vulnerable code snippet, and the latter denotes the ending of the vulnerable code snippet.
Correspondingly, the output sequence is made up of multiple correct code snippets.
Each correct code snippet is labeled with the special tags \revise{$\langle \texttt{ModStart}\rangle$} and $\langle \texttt{ModEnd}\rangle$,  where the former denotes the beginning of the correct code snippet, and the latter denotes the ending of the correct code snippet.

Unlike previous work simply taking the full vulnerable and correct functions as inputs and outputs \cite{tufano2019empirical}, such special tags make the model focus on the areas of the vulnerable code snippet.
Meanwhile, the tags support the generation of sequences shorter than a full function, helping the model to fix long vulnerable functions (discussed in Section \ref{sec:size}).

\begin{table}[t]
\centering
\caption{Number of Top-10 CWE examples in our dataset.}
\label{tab:dataset}
    \begin{tabular}{cc}
    \toprule
    \textbf{CWE ID} & \textbf{Examples in train/valid/test} \\
    \midrule
    \textbf{CWE-119} & 1402/209/386 \\
    \textbf{CWE-125} & 602/85/170 \\
    \textbf{CWE-20} & 455/64/152 \\
    \textbf{CWE-264} & 259/35/71 \\
    \textbf{CWE-476} & 253/36/70 \\
    \textbf{CWE-200} & 251/35/64 \\
    \textbf{CWE-399} & 204/32/60 \\
    \textbf{CWE-362} & 207/29/54 \\
    \textbf{CWE-787} & 200/25/53 \\
    \textbf{CWE-190} & 186/25/59 \\
    \bottomrule
    \end{tabular}%
\end{table}

\subsection{Experimental Setup}

\textbf{Implementation Details.}
All of our approaches are built based on PyTorch framework \cite{PyTorch}.
For studied pre-trained models, we use the Hugging Face~\cite{huggingface} 
implementation version of  models in our work.
There may exist two model architectures at different sizes for some pre-trained models, e.g., \texttt{CodeT5-base} and \texttt{CodeT5-large}.
According to previous work~\cite{raffel2019t5,yuan2022circle},  we utilize the base version as the initial point,  as the base version is quite practical to employ with comparable effectiveness compared to the large version.

\textbf{Training Parameters.}
We adopt the default training parameters in previous studies \cite{fu2022vulrepair,chen2022neural}.
In particular, the optimizer is Adam~\cite{kingma2014adam} with $5e-5$ learning rate.
The batch size is $8$ and we train for most $75$ epochs.
The max length of the input and output sequence is set to $512$ and $256$ due to the model limitation.
The neural beam size determines the number of generated patches and plays a crucial role in the repair inference phase.
We set the beam size to 50 in our experiment according to most repair studies \cite{fu2022vulrepair,chen2022neural}.
We will also discuss the impact of beam size in Section \ref{sec:beamsize}.

\textbf{Hardware Parameters.}
All the training and evaluation of the studied methods are conducted on one Ubuntu 18.04.3 server with two Tesla V100-SXM2 GPUs.

\section{Results and Analysis}
\label{sec:re&an}

\subsection{RQ1: Effectiveness of Pre-trained Model-based Vulnerability Repair}
\label{sec:effectiveness}

\textbf{\textit{Design.}}
In this section, we aim to explore the actual feasibility of applying large-scale pre-trained models to repair security vulnerabilities. 
In particular, we consider \delete{four}\revise{five} state-of-the-art pre-trained models (i.e., CodeT5, CodeBERT, GraphCodeBERT\delete{ and UniXcoder}\revise{, UniXcoder and CodeGPT}) as the subjects and select the most recent technique VRepair as the baseline, \delete{which has shown promising results on the vulnerability repair task}\revise{which is the most relevant technique to our work and represents state-of-the-art}.
We evaluate the effectiveness of vulnerability repair approaches using the percentage of perfect prediction accuracy.
Then, we compare the prediction accuracy of the pre-trained models with VRepair using two vulnerability datasets, summarized as follows.

\begin{itemize}
    \item \textbf{M1}: A pre-trained CodeT5 model with an Encoder-Decoder architecture for vulnerability repair.
    \item \textbf{M2}: A pre-trained CodeBERT model with an Encoder-only architecture for vulnerability repair.
    \item \textbf{M3}: A pre-trained GraphCodeBERT model with an \delete{Encoder-Decoder}\revise{Encoder-only} architecture for vulnerability repair.
    \item \textbf{M4}: A pre-trained UniXcoder model with an Encoder-Decoder architecture for vulnerability repair.
    \item \revise{\textbf{M5}: A pre-trained CodeGPT model with an Decoder-only architecture for vulnerability repair.}
    \item \textbf{M6}: A vanilla Encoder-Decoder Transformer model called VRepair for vulnerability repair.

\end{itemize}

\begin{figure}[t]
    \centering
    \includegraphics[width=0.95\linewidth]{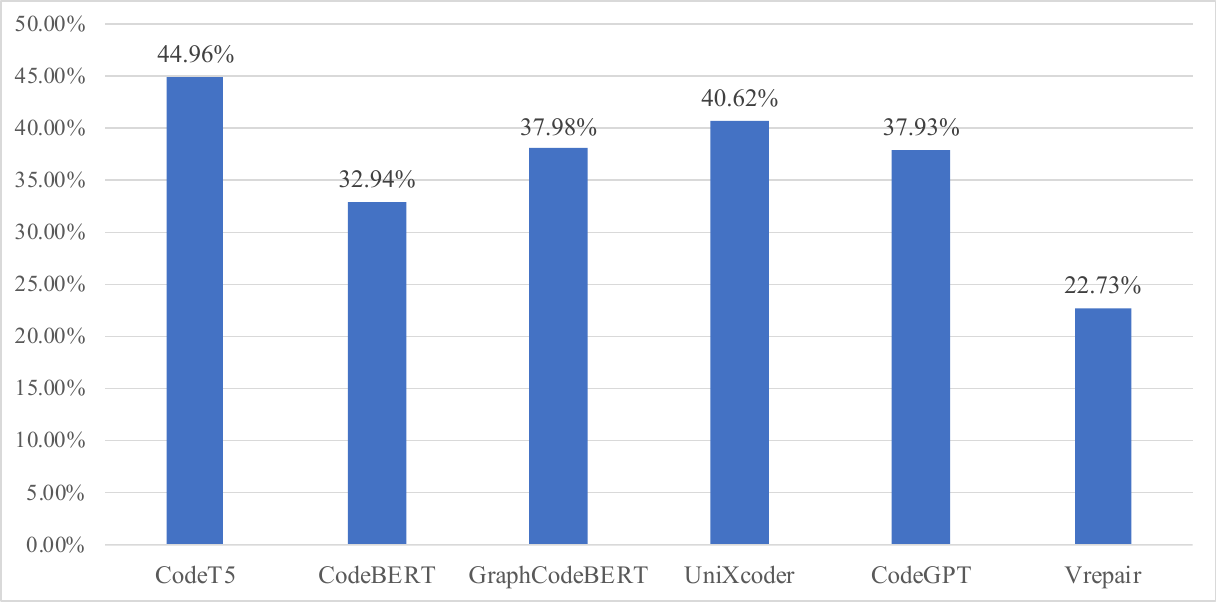}
    \caption{\revise{The experimental results of pre-trained models and VRepair in vulnerability repair.}}
    \label{fig:rq1_cve}
\end{figure}

\textbf{\textit{Results.}}
Fig. \ref{fig:rq1_cve} presents the comparison results of pre-trained models and the baseline approach VRepair in terms of prediction accuracy.
When comparing pre-trained models against VRepair, pre-trained models achieve a prediction accuracy of \delete{39.13}\revise{38.89}\% on average, which is \delete{16.40~}\revise{16.16}\% more accurate than VRepair.
\delete{
In particular, Fig. \ref{fig:rq1_cve} shows that VRepair achieves a perfect prediction of 22.73\%, while CodeT5, CodeBERT, GraphCodeBERT and UniXcoder achieve a perfect prediction of 38.92\%, 29.07\%, 33.65\% and 32.77\%, indicating that pre-trained models improve VRepair by 22.23\%, 10.21\%, 15.25\% and 17.89\%, respectively.}
\revise{
In particular, Fig. \ref{fig:rq1_cve} shows that VRepair achieves a perfect prediction of 22.73\%, while CodeT5, CodeBERT, GraphCodeBERT, UniXcoder and CodeGPT achieve a perfect prediction of 44.96\%, 32.94\%, 37.98\% 40.62\% and 37.93\%, indicating that these pre-trained models improve VRepair by 22.23\%, 10.21\%, 15.25\%, 17.89\% and 15.20\%, respectively.
}
Based on our analysis of these results, we observe that the possible reasons for the prediction accuracy improvement over VRepair lie in 
(1) \delete{pre-training~}\revise{pre-trained} models utilize large codebases to generate more meaningful vector representation (e.g., the pre-training data of UniXcoder includes 2.3 million functions from CodeSearchNet across six different programming languages), while VRepair is pre-trained on a limited \delete{bug-fix}\revise{bug fixing} corpus of 23,607 C/C++ functions; 
(2) pre-trained models utilize \delete{a Byte-Pairs Encoding (BPE)}\revise{subword-level} tokenization to handle out-of-vocabulary (OOV) issues while VRepair leverages word-level tokenization; 
(3) pre-trained models utilize T5 or BERT architecture that considers the relative position information in the self-attention mechanism while VRepair considers a vanilla transformer (i.e., a default Transformer architecture \cite{vaswani2017attention}).

When comparing the performance among different pre-trained models, we find CodeT5 achieves optimal performance.
The improvement against CodeBERT, GraphCodeBERT and UniXcoder reaches 12.02\%, 6.98\%\delete{ and }\revise{, }4.34\%\revise{, and 7.03\%}, respectively.
We observe there are two main reasons for the advance of CodeT5.
(1) CodeT5 is pre-trained on CodeSearchNet (containing Python, \delete{Javascript~}\revise{JavaScript}, Ruby, Go, Java and PHP) and BigQuery (containing one million C language functions) datasets  while the other three models are only pre-trained on CodeSearchNet dataset.
Thus, CodeT5 can acquire adequate general knowledge about C language, which is then transferred into vulnerability repair;
(2) CodeT5 is equipped with an encoder-decoder architecture, which can support generation tasks well in previous work \cite{wang2021codet5}.
However, encoder-only models (e.g., CodeBERT) require a decoder for generation tasks where the decoder initializes from scratch and cannot benefit from the pre-training dataset.

\myfinding{1}{Our comparison results reveal that,
(1) pre-trained models can achieve remarkable performance in automated software vulnerability repair with 32.94\%$\sim$44.96\% prediction accuracy;
(2) compared to the state-of-the-art technique VRepair, pre-trained models substantially achieve better repair performance and the improvement reaches 10.21\%$\sim$22.23\%;
(3) CodeT5 performs better than other studied pre-trained models by 4.34\%$\sim$12.02\% higher prediction accuracy due to the encoder-decoder architecture and pre-training dataset.
}

\begin{figure*}[htbp]
\graphicspath{{figs/}}
\centering
    \subfigure[\newrevise{Code Context}]
    {
        \includegraphics[width=0.31\linewidth]{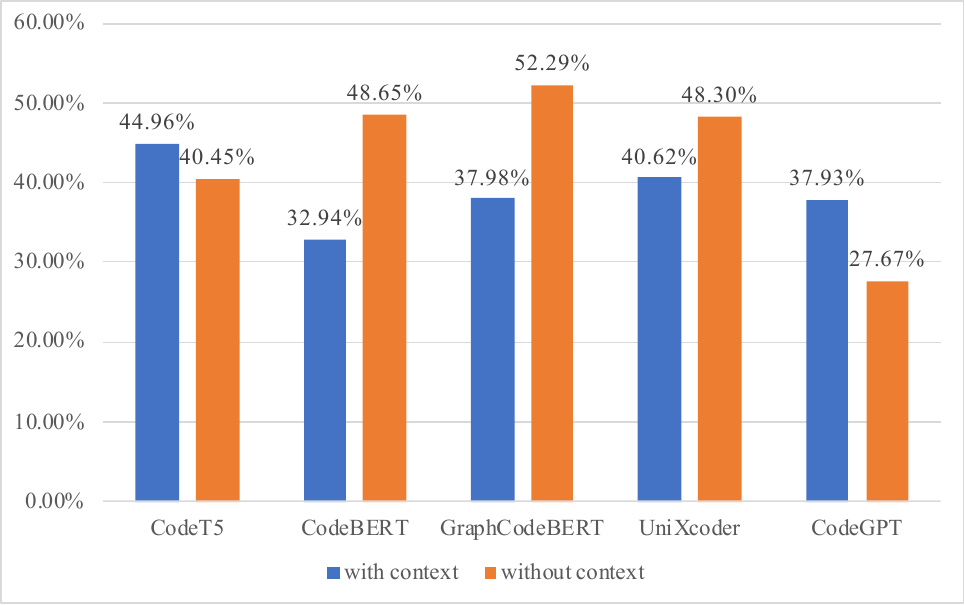}
        \label{fig:context}
    }
    \subfigure[Code Abstraction]
    {
        \includegraphics[width=0.31\linewidth]{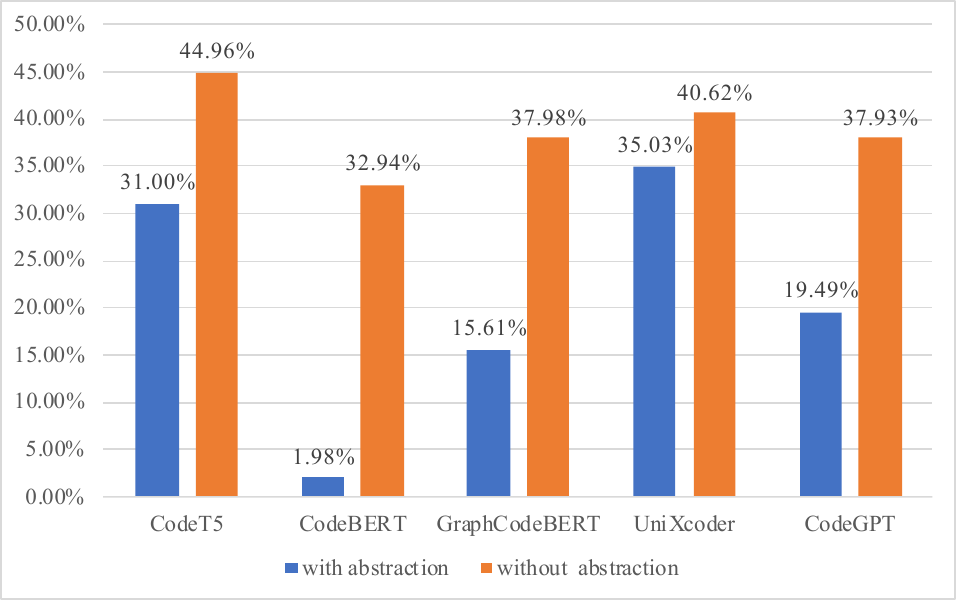}
        \label{fig:abstraction}
    }
    \subfigure[Code Tokenization]
    {
        \includegraphics[width=0.31\linewidth]{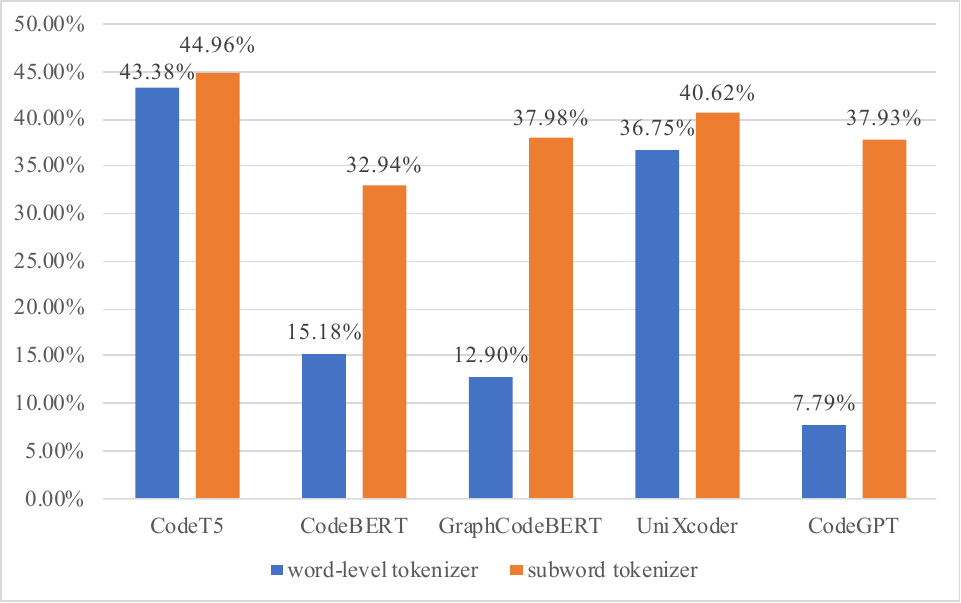}
        \label{fig:tokenizer}
        
    }
    \caption{\newrevise{The experimental results of pre-trained models in the dataset pre-processing phase.}}
    \label{fig:preprocess}
\end{figure*}

\subsection{RQ2: The Analysis in Dataset Pre-processing.}
In this section, we aim to investigate how different dataset pre-processing choices can influence the vulnerability repair performance of pre-trained models.
In particular, according to previous repair work \cite{chen2019sequencer}, our analysis mainly lies in three aspects: code context, abstraction and tokenization.

\subsubsection{RQ2.1: The Impact of Code Context}
\textbf{\emph{Design.}}
Code context generally refers to other
correct lines around the vulnerable lines in a vulnerable function.
In the manual repair scenario, the code context plays a crucial role in understanding the faulty behavior.
Security experts usually identify the vulnerable lines, and then analyze how they interact with the rest of the lines in the function in order to reason about the possible fix.
In RQ1, following VRepair \cite{chen2022neural}, we feed the full vulnerable function into the repair model as the context.
Thus, in this section, we analyze the impact of code context on the vulnerability repair and explore whether an NMT-based repair model may have diverse prediction results with or without code context.

To answer this RQ, we aim to investigate the impact of the code context for vulnerability repair.
We conduct an experiment of different \delete{pre-training~}\revise{pre-trained} models with or without code context. 
In total, we extend our experiment to systematically evaluate the following \delete{8}\revise{ten} variants of pre-trained model-based vulnerability repair approaches, i.e., two context settings (\revise{vulnerable function }with context,  \revise{vulnerable line }without context) × \delete{four}\revise{five} \delete{pre-training~}\revise{pre-trained} models (CodeT5, CodeBERT, GraphCodeBERT\delete{ and UniXcoder}\revise{, UniXcoder and CodeGPT}).

\begin{itemize}
    \item \textbf{M1-M2}: A pre-trained CodeT5 model with and without code context.
    \item \textbf{M3-M4}: A pre-trained CodeBERT model with and without code context.
    \item \textbf{M5-M6}: A pre-trained GraphCodeBERT model with and without code context.
    \item \textbf{M7-M8}: A pre-trained UniXcoder model with and without code context.
    \item \revise{\textbf{M9-M10}: A pre-trained CodeGPT model with and without code context.}
\end{itemize}

\textbf{\emph{Results.}}
The results are presented in Fig. \ref{fig:context}.
We can find that removing context from vulnerability functions improves the percentage of prediction accuracy by 8.30\% on average for pre-trained model-based vulnerability repair approaches.
In particular, CodeBERT with removing context achieves an impressive prediction accuracy of 48.65\%, which is 15.71\% higher than CodeBERT with reserving context.
Similarly, the improvement for GraphCodeBERT and UniXcoder reaches 14.31\% (52.29\%-37.98\%) and 7.68\% (48.30\%-40.62\%), respectively.
It's worth noting that GraphCodeBERT without context achieves a new record of 52.29\% prediction accuracy, which is 29.56\% higher than VRepair.
We also find the prediction accuracy of removing context decreases by 4.51\% (44.96\%-40.45\%) compared with reserving context for CodeT5, demonstrating that \delete{reserving}\revise{removing} context fails to provide sustainable benefits for all pre-trained models.
\revise{The possible reason lies in that it is difficult for encoder-only models (e.g., CodeBERT) to capture the long-term dependency relationships in code context as they are not pre-trained with C language corpora, resulting in a higher performance without code context.}
Therefore, such observation motivates security researchers to further investigate the impact of code context for specific pre-trained models in vulnerability repair.

\subsubsection{RQ2.2: The Impact of Code Abstraction}

\textbf{\emph{Design.}}
Previous work demonstrates that it is challenging for transformers to learn bug fixing transformations due to the huge vocabulary of source code \cite{tufano2019empirical}.
In particular, transformers usually employ a beam-search decoding strategy to output repair candidates by a probability distribution over all words.
The search space can be extremely large with many possible words in the source code.  
Code abstraction can limit the number of words the transformer needs to process by renaming raw words (e.g., function names and string literals) to a set of predefined tokens.
In vulnerability repair, SeqTrans \cite{chi2022seqtrans} adopts abstracted source code to make the transformer concentrate on learning common patterns from different code changes, while VRepair \cite{chen2022neural} adopts the raw source code as code abstraction may hide valuable information about the variable that can be learned by word embedding.
Thus, we investigate whether the code abstraction benefits pre-trained models in vulnerability repair.

\revise{
Tufano et al.~\cite{tufano2019empirical} release a code abstraction tool src2abs. However, src2abs cannot be employed in our experiment as it focuses on the Java language.}
We implement a tool named {\toolname}\footnote{{\toolname} is available in our replication package \delete{to facilitate}\revise{for} future studies \cite{myurl}.} to ab\textbf{S}tract \textbf{C} l\textbf{AN}guage source code based on ANother Tool for Language Recognition (ANTLR) \cite{parr2011ll}.
First, {\toolname} builds a C Lexer to process and tokenize the source code.
Second, {\toolname} builds a C Parser that analyzes the code and identifies the identifier and literal types in the source code based on Abstract Syntax Tree representation.
Third, {\toolname} replaces each identifier and literal in the stream of tokens with a unique ID which represents the type and role of the identifier/literal in the source code.
For example, function names \textit{getVal} and \textit{processRequest} can be replaced with \textit{func\_1} and \textit{func\_2}.
Finally, {\toolname} maintains a dictionary to store the mappings between the \delete{raw source code and the abstracted one}\revise{raw and abstracted source code} to be reconstructed after repair prediction.

To answer this RQ, we aim to investigate the impact of code abstraction in vulnerability repair.
We conduct an experiment of different \delete{pre-training~}\revise{pre-trained} models with or without code abstraction. 
In total, we extend our experiment to systematically evaluate the following \delete{8}\revise{ten} variants of pre-trained model-based vulnerability repair approaches, i.e., two abstraction settings (with abstracted code and with raw code) × \delete{four}\revise{five} \delete{pre-training~}\revise{pre-trained} models (CodeT5, CodeBRT, GraphCodeBERT\delete{ and UniXcoder}\revise{, UniXcoder and CodeGPT}).

\begin{itemize}
    \item \textbf{\newdelete{M1-M2}\newrevise{M11-M12}}: A pre-trained CodeT5 model with and without code abstraction.
    \item \textbf{\newdelete{M1-M2}\newrevise{M13-M14}}: A pre-trained CodeBERT model with and without code abstraction.
    \item \textbf{\newdelete{M1-M2}\newrevise{M15-M16}}: A pre-trained GraphCodeBERT model with and without code abstraction.
    \item \textbf{\newdelete{M1-M2}\newrevise{M17-M18}}: A pre-trained UniXcoder model with and without code abstraction.
    \item \revise{\textbf{\newdelete{M1-M2}\newrevise{M19-M20}}: A pre-trained CodeGPT model with and without code abstraction.}
\end{itemize}

\textbf{\emph{Results.}}
Fig. \ref{fig:abstraction} presents the experimental results of code abstraction when pre-trained models are used in vulnerability repair.
We can find that code abstraction decreases the percentage of perfect prediction by 5.59\%\delete{-}\revise{$\sim$}30.96\% for all pre-trained models in vulnerability repair.
In particular, the use of raw source code improves the prediction accuracy by 13.96\% (44.96\%-31.00\%) for CodeT5, 30.96\% (32.94\%-1.98\%) for CodeBERT, 22.37\% (37.98\%-15.61\%) for GraphCodeBERT, 5.59\% (40.62\%-35.03\%) for UniXcoder\revise{, and 18.44\% (37.93\%-19.49\%) for CodeGPT}.
This finding highlights the substantial benefits of using raw source code over abstracted source code in pre-trained model-based vulnerability repair approaches.
After our analysis, the possible reason is that code abstraction may hide valuable information about the raw works that can be learned by word embeddings.
For example, \textit{GetResult} should be a function that returns a result.
Thus, the meaningful vector representation pre-trained on a larger codebase (e.g., CodeT5 is pre-trained with 8.35 million functions from 8 different programming languages) cannot be utilized, resulting in a decreased performance in vulnerability repair.

\subsubsection{RQ2.3: The Impact of Code Tokenization}

\textbf{\emph{Design.}}
Code tokenization converts the source code into a stream of tokens, which are in turn used by the NMT-based model for further processing and training.
Specifically, there exist two main granularities of code tokenizers used in NMT-based vulnerability repair: word-level tokenizers \cite{fu2022vulrepair} and subword-level tokenizers \cite{chi2022seqtrans}.
The word-level tokenization means that a sentence is divided according to its words (e.g., space-separated).
However, words (e.g., variable names) in programming languages can be created arbitrarily, leading to an excessively large vocabulary size.
The subword-level tokenization can reduce the vocabulary size by splitting rare tokens into multiple subwords instead of adding the full tokens into the vocabulary directly.

To answer this RQ, we aim to investigate the impact of the tokenization component in vulnerability repair. 
In total, we extend our experiment to systematically evaluate the following \delete{eight}\revise{ten} variants of pre-trained model-based vulnerability repair approaches, i.e., 2 tokenizers (subword-level \delete{tokenizer }and word-level tokenizer) × \delete{4 pre-trained models (CodeT5, CodeBERT, GraphCodeBERT, UniXcoder)}\revise{5 pre-trained models (CodeT5, CodeBERT, GraphCodeBERT, UniXcoder and CodeGPT)}.

\begin{itemize}
    \item \textbf{\newdelete{M1-M2}\newrevise{M21-M22}}: A pre-trained CodeT5 model with a subword-level and word-level tokenizer.
    \item \textbf{\newdelete{M1-M2}\newrevise{M23-M24}}: A pre-trained CodeBERT model with a subword-level and word-level tokenizer.
    \item \textbf{\newdelete{M1-M2}\newrevise{M25-M26}}: A pre-trained GraphCodeBERT model with a subword-level and word-level tokenizer.
    \item \textbf{\newdelete{M1-M2}\newrevise{M27-M28}}: A pre-trained UniXcoder model with a subword-level and word-level tokenizer.
    \item \revise{\textbf{\newdelete{M1-M2}\newrevise{M29-M30}}: A pre-trained CodeGPT model with a subword-level and word-level tokenizer.}
\end{itemize}

\textbf{\emph{Results.}}
Fig. \ref{fig:tokenizer} presents the comparison results of word-level and subword-level tokenization for vulnerability repair in terms of prediction accuracy. 
We find that the subword-level tokenization improves the percentage of perfect prediction by 1.58\%\delete{-}\revise{$\sim$}\delete{25.08}\revise{30.14}\% against word-level tokenization for all pre-trained model approaches.
\delete{
In particular, the use of subword-level tokenization improves perfect prediction accuracy by 1.58\% (44.96\%-43.38\%) for CodeT5, 17.76\% (32.94\%-15.18\%) for CodeBERT, 25.08\% (36.75\%-12.90\%) for GraphCodeBERT and 3.87\% (40.62\%-37.98\%) for UniXcoder.}
\revise{In particular, the use of subword-level tokenization improves perfect prediction accuracy by 1.58\%~(44.96\%-43.38\%) for CodeT5, 17.76\%~(32.94\%-15.18\%) for CodeBERT, 25.08\%~(36.75\%-12.90\%) for GraphCodeBERT, 3.87\%~(40.62\%-37.98\%) for UniXcoder, and 30.14\%~(37.93\%-7.79\%) for CodeGPT.}
Compared with the word-level tokenization, the subword-level tokenization can process rare words that are not even in the training data while maintaining a manageable vocabulary size.
These results highlight the substantial benefits of using subword-level tokenization for pre-trained model-based vulnerability repair approaches.

\myfinding{2}{The performance under different pre-processing approaches demonstrates that:
(1) removing code context improves the prediction accuracy by \delete{8.30}\revise{4.07}\% on average;
(2) code abstraction hinders the utilization of generic knowledge in pre-trained models for vulnerability repair, resulting in an average of \delete{18.22}\revise{18.26}\% decrease in prediction accuracy;
(3) subword-level tokenization improves word-level tokenization with \delete{9\%$\sim$14\%}\revise{15.69\%} higher prediction accuracy\revise{ on average} for vulnerability repair\delete{ approaches}.
}

\subsection{RQ3: The Analysis in Model Training.}
\label{sec:rq3}

In this section, we aim to investigate how different components in the training phase can influence the vulnerability repair performance of pre-trained models.
In particular, we explore the contribution of each training component (i.e., pre-training and fine-tuning) by examining the prediction accuracy of pre-trained models when equipped with each component separately.

\subsubsection{RQ3.1: The Role of Pre-training}

\textbf{\emph{Design.}}
We have demonstrated that pre-trained models achieve better prediction accuracy than the baseline VRepair in Section \ref{sec:effectiveness}.
We notice that VRepair is trained in a traditional pipeline (i.e., supervised learning on label datasets).
However, the model training process of pre-trained models is two-fold: 
(1) a pre-training process for a general task (e.g., next word prediction) with unsupervised learning on large unlabeled datasets;
and (2) a fine-tuning process for a specific task (e.g., vulnerability repair) with supervised learning on limited labeled datasets.
It is crucial to explore the practical role of these two components separately when pre-trained models are applied to vulnerability repair.

Thus, we aim to investigate the impact of the pre-training component in vulnerability repair.
Considering that there exists \delete{two}\revise{three} model architectures (i.e., BERT with encoder-only\delete{ and}\revise{,} T5 with encoder-decoder\revise{, and GPT with decoder-only}) across the \delete{four}\revise{five} \newrevise{investigated} pre-trained models, we conduct an experiment of the pre-training component with the \delete{T5 and BERT}\revise{BERT, T5 and GPT} architectures. 
In total, we extend our experiment to systematically evaluate the following \delete{four}\revise{six} variants of pre-trained model-based vulnerability repair approaches, i.e., 2 pre-training components (pre-training and \newdelete{no}\newrevise{without} pre-training) × \delete{2 model architectures (T5 and BERT)}\revise{3 model architectures (T5, BERT and GPT)}.

\begin{itemize}
    \item \textbf{M1}: 
    A T5 Transformer architecture with pre-training, \newrevise{i.e., a pre-trained CodeT5}.
    
    \item \textbf{M2}: 
    An original T5 Transformer architecture without pre-training,\newrevise{i.e., a no-pre-trained CodeT5}.
    
    \item \textbf{M3}: 
    A BERT Transformer architecture with pre-training, \newrevise{i.e., a pre-trained CodeBERT}.
    
    \item \textbf{M4}: 
    An original BERT Transformer architecture without pre-training, \newrevise{i.e., a no-pre-trained CodeBERT}.
    
    \item \textbf{M5}: 
    A GPT Transformer architecture with pre-training, \newrevise{i.e., a pre-trained CodeGPT}.
    
    \item \textbf{M6}: 
    An original GPT Transformer architecture without pre-training, \newrevise{i.e., a no-pre-trained CodeGPT.}
 \end{itemize}

\begin{figure}[t]
    \centering
    \includegraphics[width=0.9\linewidth]{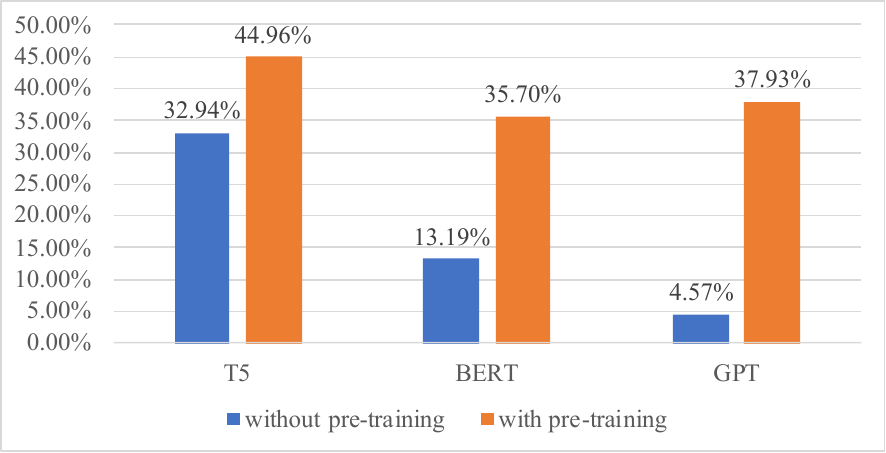}
    \caption{\newrevise{The experimental results of \newdelete{pre-trained models}\newrevise{three model architectures} with and without pre-training in vulnerability repair.}}
    \label{fig:pre-training}
\end{figure}

\textbf{\emph{Results.}}
Fig. \ref{fig:pre-training} presents the experimental results of the benefits of using pre-training components for vulnerability repair.
Regardless of the model architectures, the pre-training component improves the percentage of perfect prediction by \delete{12.02\%-22.51\%}\revise{12.02\%$\sim$33.36\%} for vulnerability repair approaches.
In particular, we can find that the pre-training corpus improves perfect prediction accuracy by 12.02\% (44.96\% - 32.94\%) for the T5 architecture\delete{ and}\revise{,} 22.51\% (35.70\% - 13.19\%) for the BERT architecture\revise{, and 31.33\% (37.93\%-4.57\%) for GPT architecture} when they are not pre-trained on any corpus (i.e., the original T5/BERT/GPT architecture provided by the Transformer library \cite{PyTorch}). 
This finding highlights the substantial benefits of the pre-training process on the larger codebase, such as CodeSearchNet with 7.35 million functions from 6 different programming languages (i.e., Ruby, JavaScript, Go, Python, Java, PHP) for vulnerability repair. 
We also find that the T5 architecture outperforms the \delete{BERT architecture}\revise{BERT and GPT architectures} when using the same pre-training components.
\revise{For example, }Fig. \ref{fig:pre-training} shows that, the T5 architecture outperforms the BERT architecture by 19.75\% (32.94\%- 13.19\%) without pre-training and by 9.26\% (44.96\%-35.70\%) with pre-training, highlighting the substantial benefits of the encoder-decoder transformer (i.e., T5) for the vulnerability repair task over the encoder-only Transformer (i.e., BERT), which could be more suitable for other code-related tasks such as code classification.

\subsubsection{RQ3.2: The \delete{Role}\revise{Impact} of Fine-Tuning \revise{Data Size}}

\textbf{\emph{Design.}}
In this RQ, we aim to investigate how the size of the training data for fine-tuning pre-trained models influences their performance in vulnerability repair.
In particular, we plot prediction accuracy curves by varying the amount of training data used to fine-tune pre-trained models.
We select 0.00\%$\sim$100\% data from the training set with 20\% intervals each time for training, and then test on the same entire testing set.
In total, we extend our experiment to systematically evaluate the following \delete{24}\revise{30} variants of pre-trained model-based vulnerability repair approaches, i.e., 6 fine-tuning components (0.00\%, 20.00\%, 40.00\%, 60.00\%, 80.00\% and 100.00\% fine-tuning corpus) × \delete{4}\revise{5} \newdelete{model architectures}\newrevise{pre-trained models} (CodeT5, CodeBERT, GraphCodeBERT\delete{ and UniXcoder}\revise{, UniXcoder and CodeGPT}).

\begin{itemize}
    \item \textbf{M1-M6}: A pre-trained CodeT5 model with different fine-tuning corpora.
    \item \textbf{M7-M12}: A pre-trained CodeBERT model with different fine-tuning corpora.
    \item \textbf{M13-M18}: A pre-trained GraphCodeBERT model with different fine-tuning corpora.
    \item \textbf{M19-M24}: A pre-trained UniXcoder model with different fine-tuning corpora.
    \item \revise{\textbf{M25-M30}: A pre-trained CodeGPT model with different fine-tuning corpora.}
\end{itemize}

\begin{figure}[t]
\graphicspath{{figs/}}
\centering
    \subfigure[CodeT5]
    {
        \includegraphics[width=0.47\linewidth]{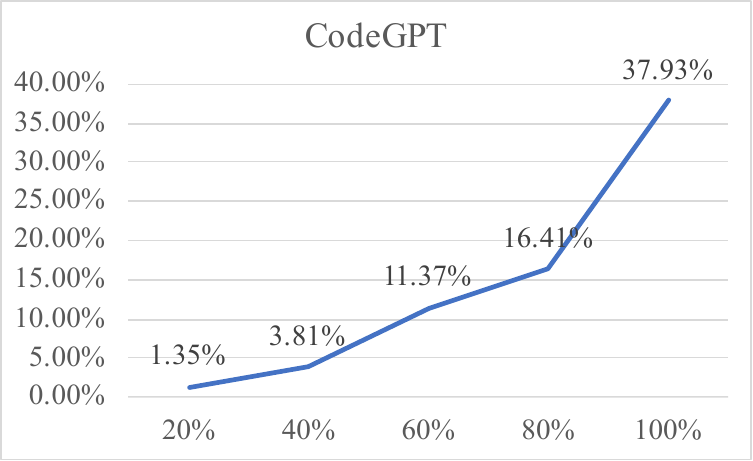}
    }
    \subfigure[CodeBERT]
    {
        \includegraphics[width=0.47\linewidth]{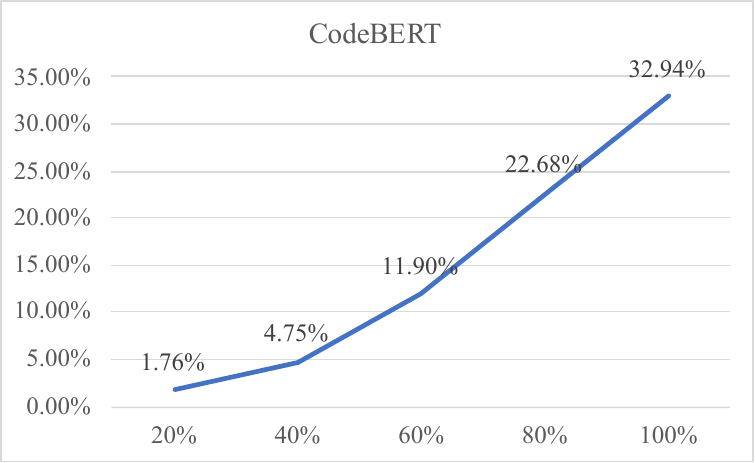}
    }
    \subfigure[GraphCodeBERT]
    {
        \includegraphics[width=0.47\linewidth]{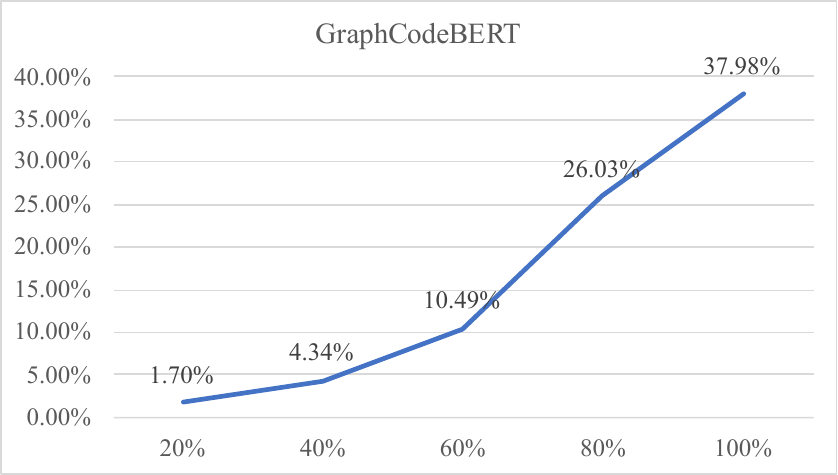}
    }
    \subfigure[UniXcoder]
    {
        \includegraphics[width=0.47\linewidth]{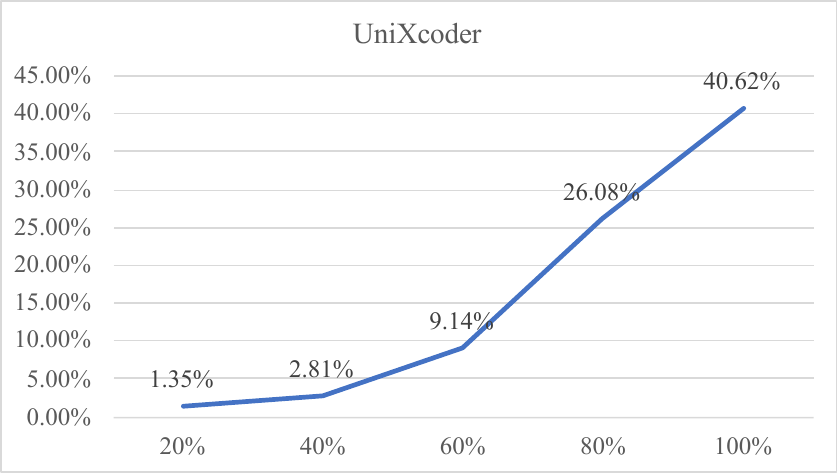}
    }
    \subfigure[CodeGPT]
    {
        \includegraphics[width=0.47\linewidth]{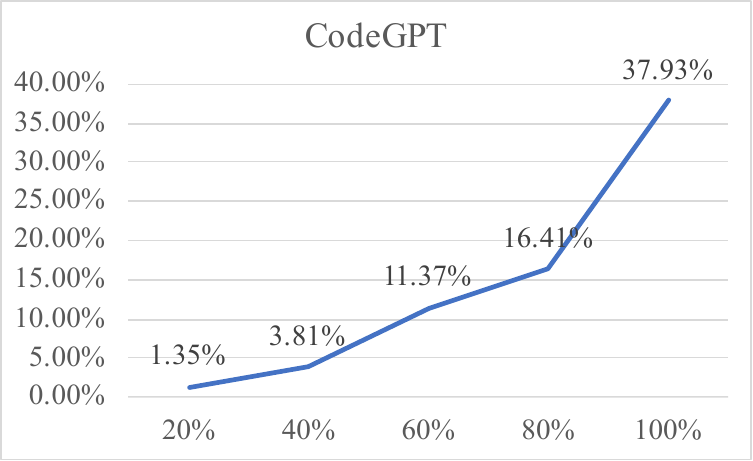}
    }
    \caption{\revise{The experimental results of pre-trained models with different fine-tuning components in vulnerability repair.}}
    \label{fig:fine-tuning}
\end{figure}

\textbf{\emph{Results.}}
Table \ref{fig:fine-tuning} presents the experimental results of fine-tuning components for vulnerability repair.
It can be seen that the performance shows an increasing trend in prediction accuracy as the size of the training set increases. 
It proves that even for pre-trained models, the size of the training set plays a key role in the performance as well.
Meanwhile, considering Fig. \ref{fig:rq1_cve}, we see that when the training set size is reduced to 80\% of the original size, the results of pre-trained models are comparable to or even higher than that of VRepair.
Further, we find that the prediction accuracy of pre-trained models increases at a nearly linear rate.
Even with all the data, there is still no trend to slow down in the growth of prediction accuracy.
This indicates that the repair performance has not yet reached saturation and will continue to increase as the data set grows.
Future research can be conducted to explore the performance of pre-trained models with a richer training set.

We also find that all pre-trained models without vulnerability repair fine-tuning cannot fix any vulnerable function.
Based on our analysis of these results, we observe that the possible reason lies in that,  although these models contain valuable knowledge from pre-training stages, they have not been adapted to the downstream task.
As a result, these models have no idea about the output format for vulnerability repair.
The results demonstrate the benefit of fine-tuning stages for vulnerability repair,  which not only allows the model to learn the task-specific knowledge, but also helps to make better use of the pre-training knowledge.
Future research can be conducted to explore the direct usage of pre-trained models\delete{ (with no fine-tuning)} for vulnerability \revise{repair}\revise{ in a zero-shot setting}.
For example, we can replace the vulnerable line with a mask line and query the models to fill the mask line with replacement tokens to produce candidate patches.

\myfinding{3}{The performance under different training components demonstrates that:
(1) the pre-training component effectively acquires general knowledge about programming languages, so as to facilitate the vulnerability repair task;
(2) the prediction accuracy shows a decreasing trend as fine-tuning datasets decrease, proving the fine-tuning component plays a key role in the vulnerability repair task.
}

\begin{figure}[t]
    \centering
    \includegraphics[width=0.95\linewidth]{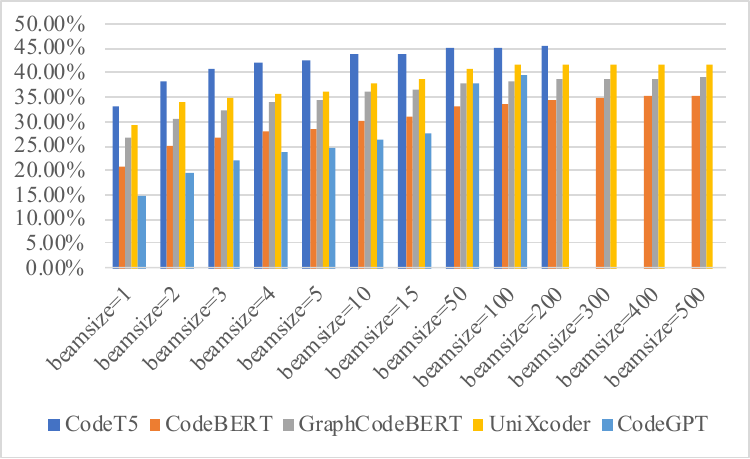}
    \caption{\revise{The experimental results of pre-trained models with different values of beam size in vulnerability repair.}}
    \label{fig:beam}
\end{figure}

\subsection{RQ4: The Analysis in Repair Inference}

In this section, we aim to investigate how pre-trained models perform with different choices in the inference phase for vulnerability repair.
In particular, we focus on beam size and vulnerable function size under repair.

\subsubsection{RQ4.1: The Impact of Beam Size}
\label{sec:beamsize}

\begin{table*}[htbp]
\centering
\caption{\revise{The comparison results of the four pre-trained models when repairing vulnerabilities with different lengths of tokens. 
The first and second column list the number of tokens and the remaining columns list the corresponding prediction accuracy.
We use the shade of the color to represent the value in the cell, where dark green denotes the highest accuracy values, and dark red for the lowest accuracy values.
}}
\label{tab:size}
\renewcommand\arraystretch{1.3}
\begin{tabular}{c|c|ccccc|c}
\hline
    \multicolumn{2}{c|}{\textbf{\#Tokens}} & \textbf{CodeT5} & \textbf{CodeBERT} & \textbf{GraphCodeBERT} & \textbf{UniXcoder} & \revise{\textbf{CodeGPT}} & \revise{\textbf{Average}} \\
    \hline
    \multirow{6}[4]{*}{\textbf{$\leq$500}} & \textbf{0-100} & \cellcolor[rgb]{0.388, 0.745, 0.482} 64.91\% & \cellcolor[rgb]{0.388, 0.745, 0.482} 55.14\% & \cellcolor[rgb]{0.388, 0.745, 0.482} 53.27\% & \cellcolor[rgb]{0.388, 0.745, 0.482} 55.81\% & \cellcolor[rgb]{0.388, 0.745, 0.482} \revise{57.69\%} & \cellcolor[rgb]{0.388, 0.745, 0.482} \revise{57.36\%} \\
          & \textbf{100-200} & \cellcolor[rgb]{0.710, 0.839, 0.502} 59.85\% & \cellcolor[rgb]{0.906, 0.894, 0.514} 38.82\% & \cellcolor[rgb]{0.710, 0.839, 0.502} 47.45\% & \cellcolor[rgb]{0.780, 0.859, 0.506} 47.79\% & \cellcolor[rgb]{0.745, 0.851, 0.506} \revise{49.80\%} & \cellcolor[rgb]{0.792, 0.863, 0.506} \revise{48.74\%} \\
          & \textbf{200-300} & \cellcolor[rgb]{0.996, 0.886, 0.510} 53.66\% & \cellcolor[rgb]{0.996, 0.910, 0.514} 35.64\% & \cellcolor[rgb]{0.996, 0.871, 0.506} 41.09\% & \cellcolor[rgb]{0.961, 0.914, 0.518} 44.08\% & \cellcolor[rgb]{0.973, 0.914, 0.518} \revise{44.72\%} & \cellcolor[rgb]{0.996, 0.902, 0.514} \revise{43.84\%} \\
          & \textbf{300-400} & \cellcolor[rgb]{0.804, 0.867, 0.510} 58.38\% & \cellcolor[rgb]{0.996, 0.922, 0.518} 35.96\% & \cellcolor[rgb]{0.941, 0.906, 0.518} 43.26\% & \cellcolor[rgb]{0.996, 0.871, 0.506} 42.42\% & \cellcolor[rgb]{0.996, 0.898, 0.510} \revise{43.41\%} & \cellcolor[rgb]{0.980, 0.918, 0.518} \revise{44.69\%} \\
          & \textbf{400-500} & \cellcolor[rgb]{0.988, 0.761, 0.486} 47.71\% & \cellcolor[rgb]{0.996, 0.906, 0.514} 35.57\% & \cellcolor[rgb]{0.988, 0.741, 0.482} 38.26\% & \cellcolor[rgb]{0.976, 0.522, 0.439} 36.62\% & \cellcolor[rgb]{0.980, 0.584, 0.451} \revise{34.46\%} & \cellcolor[rgb]{0.988, 0.706, 0.475} \revise{38.52\%} \\
\cline{2-8}          & \textbf{Average} & \cellcolor[rgb]{0.902, 0.894, 0.514} 56.80\% & \cellcolor[rgb]{0.902, 0.894, 0.514} 38.95\% & \cellcolor[rgb]{0.882, 0.890, 0.514} 44.33\% & \cellcolor[rgb]{0.902, 0.894, 0.514} 45.27\% & \cellcolor[rgb]{0.929, 0.902, 0.514} \revise{45.71\%} & \cellcolor[rgb]{0.910, 0.898, 0.514} \revise{46.21\%} \\
    \hline
    \multicolumn{2}{c|}{\textbf{$>$500}} & \cellcolor[rgb]{0.973, 0.412, 0.420} 30.90\% & \cellcolor[rgb]{0.973, 0.412, 0.420} 26.38\% & \cellcolor[rgb]{0.973, 0.412, 0.420} 31.04\% & \cellcolor[rgb]{0.973, 0.412, 0.420} 34.75\% & \cellcolor[rgb]{0.973, 0.412, 0.420} \revise{29.51\%} & \cellcolor[rgb]{0.973, 0.412, 0.420} \revise{30.52\%} \\
    \hline
    \multicolumn{2}{c|}{\textbf{All}} & \cellcolor[rgb]{0.988, 0.706, 0.475} 44.96\% & \cellcolor[rgb]{0.988, 0.765, 0.486} 32.94\% & \cellcolor[rgb]{0.988, 0.729, 0.478} 37.98\% & \cellcolor[rgb]{0.988, 0.761, 0.486} 40.62\% & \cellcolor[rgb]{0.988, 0.706, 0.475} \revise{37.93\%} & \cellcolor[rgb]{0.988, 0.722, 0.478} \revise{38.89\%} \\
    \hline
    
    \end{tabular}%
\end{table*}

\textbf{\emph{Design.}}
Following existing NMT-based vulnerability repair work \cite{chen2022neural}, we report the performance of pre-trained models with a beam size $k = 50$ that returns 50 possible repair candidates.
However, security researchers may spend a considerable amount of effort to assess the correctness of the generated repair candidates manually~\cite{zhang2023boosting}.
In such a scenario, only inspecting fewer repair candidates (e.g., Top-1 and Top-5) that have a high probability of being correct is more practical and reduces the valuable manual effort.
On the contrary,  when automated inspection is available (e.g., executing the test suite against the candidates), security researchers can filter the incorrect repair candidates automatically without manual inspection.
In the related bug fixing task, hundreds or even thousands of repair candidates (e.g., 1000 candidate patches per bug in CIRCLE \cite{yuan2022circle}) are generated.
In such a scenario, if a higher beam size value is set, security researchers can check more repair candidates and reasonably achieve higher prediction accuracy, which may benefit its adoption in practice.

Thus, in this section, we investigate the prediction accuracy of pre-trained models with different beam size values, so as to explore how they perform in different practical scenarios.
In total, we extend our experiment to systematically evaluate the following \delete{52}\revise{65} variants of pre-trained model-based vulnerability repair approaches, i.e., 13 beam value settings (Top-1$\sim$Top-5, Top-10, Top-15, Top-50, Top-100, Top-200, Top-300, Top-400, Top-500) × \delete{4}\revise{5} \newdelete{model architectures}\newrevise{pre-trained models} (CodeT5, CodeBERT, GraphCodeBERT\delete{ and UniXcoder}\revise{, UniXcoder and CodeGPT}).

\begin{itemize}
    \item \textbf{M1-M13}: A pre-trained CodeT5 model with different beam size values.
    \item \textbf{M14-M26}: A pre-trained CodeBERT model with different beam size values.
    \item \textbf{M27-M39}: A pre-trained GraphCodeBERT model with different beam size values.
    \item \textbf{M40-M52}: A pre-trained UniXcoder model with different beam size values.
    \item \revise{\textbf{M53-M65}: A pre-trained CodeGPT model with different beam size values.}
\end{itemize}

\textbf{\emph{Results.}}
Fig. \ref{fig:beam} presents the prediction accuracy of pre-trained models under different values of beam size\footnote{We only report the results of CodeT5 with a maximum 200 beam size \revise{and CodeGPT with a maximum 100 size} due to our devices’ limits.}.
We can find that, \revise{except for CodeGPT, } pre-trained models still achieve a prediction accuracy of 20.63\%$\sim$33.00\% when analyzing the \text{Top}-1 repair candidate (i.e., beam size $k = 1$), which is competitive against VRepair.
When the beam size value increases to 5, the prediction accuracy is improved by 7.68\%$\sim$9.73\% for all pre-trained models.   
In other words, security researchers can correctly repair 483$\sim$723 security vulnerabilities out of 1,706 testing instances when only manually inspecting the \text{Top}-5 repair candidates recommended by pre-trained models.
Meanwhile, regardless of the pre-trained models, we still see an improved Top-k prediction accuracy if we set a larger beam size value.
For example, the prediction accuracy of UniXcoder increases from 28.31\% to 35.29\% when the $k$ repair candidates are from 5 to 500, resulting in 119 additional correct vulnerability fixes.
Such improvement is valuable if security researchers can assess the repair candidate correctness automatically, highlighting the substantial benefits of a larger beam size.

\subsubsection{RQ4.2: The Impact of \delete{Vulnerability Size~}\revise{Function Length}}
\label{sec:size}

\textbf{\emph{Design.}}
The size of the input sequences is an essential part in the  repair inference phase for pre-trained models, as longer input sequences usually result in more vocabulary and larger search space.
We have demonstrated that pre-trained models can generate a considerable number of correct patches ( e.g., UniXcoder fixed 693 out of 1,706 security vulnerabilities).
However, there is a large number of 1,013 vulnerable functions that cannot be correctly repaired.
Thus, we perform a further investigation to analyze the prediction accuracy of pre-trained models (\textbf{\delete{M1-M4}\revise{M1-M5}} in Section \ref{sec:effectiveness}) with respect to the vulnerable function length.

\textbf{\emph{Results.}}
The results are presented in Table \ref{tab:size}.
It can be found that the repair performance of pre-trained models generally decreases as the size of the vulnerable functions increases.
\delete{
For example, when repairing vulnerable functions with less than 100 tokens, CodeT5, CodeBERT, GraphCodeBERT and UniXcoder have a prediction accuracy of 64.91\%, 55.14\%, 53.27\% and 55.81\%, which are 5.06\%, 16.32\%, 5.82\% and 8.02\% higher than the results over vulnerable functions with 100$\sim$200 tokens, respectively.}
\revise{For example, when repairing vulnerable functions with less than 100 tokens, CodeT5, CodeBERT, GraphCodeBERT, UniXcoder and CodeGPT have a prediction accuracy of 64.91\%, 55.14\%, 53.27\%, 55.81\% and 57.69\%, which are 5.06\%, 16.32\%, 5.82\%, 8.02\% and 7.89\% higher than the results over vulnerable functions with 100$\sim$200 tokens, respectively.}
Similarly, when repairing vulnerable functions with 100$\sim$200 tokens, the prediction accuracy is \delete{4.86}\revise{4.90}\% higher than that of vulnerable functions with \delete{100$\sim$200}\revise{200$\sim$300} tokens on average.

We also find that pre-trained models are most accurate for vulnerable functions that have less than 500 tokens.
For example, CodeT5 achieves a perfect prediction accuracy of 56.80\% on average for vulnerable functions with less than 500 tokens, but the perfect prediction substantially decreases to 30.90\% for vulnerable functions with greater than 500 tokens.
Similar performance can be seen for other models.
\delete{
The drop in prediction accuracy reaches 25.90\% (56.80\%-30.90\%), 12.57\% (38.95\%-26.38\%), 13.29\% (44.33\%-31.04\%) and 10.52\% (45.27\%-34.75\%), respectively.}
\revise{
The drop in prediction accuracy for other models reaches 25.90\% (56.80\%-30.90\%), 12.57\% (38.95\%-26.38\%), 13.29\% (44.33\%-31.04\%), 10.52\% (45.27\%-34.75\%) and 16.20\%(45.71\%-29.51\%), respectively.
}
We observe the performance decrease for large functions (500+ tokens) is mainly due to 
(1) it is difficult for neural networks to learn the correct code changes for complex vulnerable functions~\cite{yuan2022circle, zhu2021syntax};
and (2) the transformer architecture limits the maximum input sequence to 512 tokens.
As a result, for any vulnerable functions with greater than 512 tokens, such extra tokens will be truncated and will not be processed and learned by the models, leading to a negative impact on the accuracy of prediction results.
Thus, future researchers should further explore techniques that can handle larger functions (i.e., the functions with more than 512 tokens).

\begin{table*}[htbp]
  \centering
  \caption{\revise{The performance of pre-trained models on the Top-10 most common CWE types. 
The first and second columns list the Top-10 CWE types sorted by the number of vulnerabilities.
The third column list the related CWE definitions.
The remaining columns list the prediction accuracy of four pre-trained models and their average accuracy.
We also present the overall results over the Top-10 most common CWE types in the last row.} \newrevise{CT5, CBERT, GCBERT, UCoder and CGPT denote CodeT5, CodeBERT, GraphCodeBERT, UniXCoder and CodeGPT, respectively.}}
  \label{tab:cwe_type}
    \begin{tabular}{c|c|c|c|ccccc|c}
    
    \toprule
    \textbf{Rank} & \textbf{CWE Type} & \textbf{Name} & \textbf{Number} & \textbf{CT5} & \textbf{CBERT} & \textbf{GCBERT} & \textbf{UCoder} & \textbf{\revise{CGPT}} & \textbf{\revise{Average}} \\
    \toprule

    \textbf{1} & \textbf{CWE-119} & Improper Restriction of Operations & 386   & 38.08\% & 24.61\% & 28.76\% & 30.57\% & \revise{28.76\%} & \revise{30.16\%} \\
    \textbf{2} & \textbf{CWE-125} & Out-of-bounds Read & 170   & 36.47\% & 31.76\% & 35.29\% & 38.82\% & \revise{35.29\%} & \revise{35.53\%} \\
    \textbf{3} & \textbf{CWE-20} & Improper Input Validation & 152   & 50.00\% & 37.50\% & 40.13\% & 42.11\% & \revise{42.11\%} & \revise{42.37\%} \\
    \textbf{4} & \textbf{CWE-264} & Permissions, Privileges, Access Controls & 71    & 54.93\% & 43.66\% & 50.70\% & 54.93\% & \revise{44.63\%} & \revise{49.77\%} \\
    \textbf{5} & \textbf{CWE-476} & NULL Pointer Dereference & 70    & 55.71\% & 35.71\% & 45.71\% & 54.29\% & \revise{50.70\%} & \revise{48.42\%} \\
    \textbf{6} & \textbf{CWE-200} & Exposure of Sensitive Information & 64    & 71.88\% & 48.44\% & 62.50\% & 65.63\% & \revise{47.14\%} & \revise{59.12\%} \\
    \textbf{7} & \textbf{CWE-399} & Resource Management Errors & 60    & 58.33\% & 48.33\% & 50.00\% & 51.67\% & \revise{60.94\%} & \revise{53.85\%} \\
    \textbf{8} & \textbf{CWE-190} & Integer Overflow or Wraparound & 59    & 52.54\% & 32.20\% & 28.81\% & 35.59\% & \revise{50.00\%} & \revise{39.83\%} \\
    \textbf{9} & \textbf{CWE-362} & Race Condition & 54    & 46.30\% & 37.04\% & 48.15\% & 48.15\% & \revise{33.90\%} & \revise{42.71\%} \\
    \textbf{10} & \textbf{CWE-787} & Out-of-bounds Write & 53    & 28.30\% & 20.75\% & 24.53\% & 30.19\% & \revise{49.09\%} & \revise{30.57\%} \\
    \midrule
    \multicolumn{3}{c|}{\textbf{Total}} & 1141  & 45.14\% & 32.60\% & 37.34\% & 40.40\% & \revise{41.54\%} & \revise{39.40\%} \\
    \bottomrule
    
    \end{tabular}%
  \label{tab:addlabel}%
\end{table*}%

\subsubsection{\revise{The Impact of CWE Types}}

\textbf{\emph{\revise{Design.}}}
\revise{
Different types of vulnerabilities may exhibit varying debugging efforts, as some vulnerabilities might prove more challenging to repair.
The CWE denotes a list of vulnerability weakness types that can lead to a particular security issue, helping developers and security researchers gain insight into the severe security issues in their software systems.
In particular, to understand the significance of pre-trained models on the practical usage scenarios of vulnerability repair, we attempt to answer whether pre-trained models have better performance for a specific type of CWE.
We consider \textbf{M1-M5} models in Section \ref{sec:effectiveness} and perform an investigation to better understand the performance of the Top-10 majority of CWE types in the repair inference phase.
}

\textbf{\emph{\revise{Results.}}}
\revise{
The results are presented in Table \ref{tab:cwe_type}.
From Table \ref{tab:cwe_type}, we find that the investigated pre-trained models correctly repair 45.14\%, 32.60\%, 37.34\%, 40.40\% and 41.54\% of the vulnerable functions over the Top-10 most common CWEs on average.
When considering the most accurate fixes, we find a large overlap in the CWEs repaired by the different models.
For example, CWE-200 (Exposure of Sensitive Information) has the highest perfect prediction (i.e., 71.88\%, 48.44\%, 62.50\%, 65.63\%, and 47.14\%) for all pre-trained models and CWE-399 (Resource Management Errors) reaches a perfect prediction of more than 50\% for four pre-trained models (i.e., CodeT5, GraphCodeBERT, UniXCoder, and CodeGPT).
Similar performance can be observed for the most inaccurate fixes.
For example, CWE-787 (Out-of-bounds Write) is the second most difficult to fix for all pre-trained models with only 30.57\% prediction accuracy on average.
We think the occurrence is closely related to the distribution of training data.
This indicates there exists an imbalance issue of CWEs in existing training data.
We highly recommend the researchers to conduct thorough evaluations for vulnerability repair imbalance issues and explore how repair models perform under different CWEs.
}

\revise{
Thus, we further analyze the perfect prediction accuracy according to the majority of the CWEs in the dataset.
We find that, except CodeGPT, CWE-200 achieves the highest 62.11\% perfect prediction with 64 samples, while CWE-787 achieves only 25.95\% perfect prediction with 53 samples, indicating that pre-trained models still cannot accurately repair some types of rare vulnerabilities.
We also find there exist some CWE types with a high sample size that pre-trained models cannot generate correct patches.
For example, although CWE-119 has 386 samples, the perfect prediction is only 30.16\%.
That means the CWE types that achieve remarkable perfect prediction may not necessarily be the majority of CWEs in the dataset.
Thus, future researchers should further explore specific repair techniques to target security vulnerabilities that are difficult to fix.
}

 \myfinding{4}{The performance under different inference settings demonstrates that:
 (1) in a manual inspection scenario, security experts fix comparable vulnerabilities by inspecting a few repair candidates,
 e.g., CodeT5 achieves 35.99\% prediction accuracy with Top-5 candidates;
 (2) in an automated inspection scenario, security experts accomplish more correct vulnerability fixes automatically,
 e.g., 119 additional vulnerabilities are fixed with candidates from 5 to 500;
 (3) the prediction accuracy of pre-trained models depends on the size of the vulnerabilities,
 e.g., CodeT5 achieves an impressive 56.80\% prediction accuracy when considering vulnerabilities with less than 500 tokens. 
 (4) pre-trained models perform well in patching some real-world security vulnerabilities, especially  CWE-200 and CWE-399, while it is challenging to repair specific types of vulnerability like CWE -119.
}

\subsection{RQ5: An Enhanced Vulnerability Repair Approach}
\label{sec:rq5} 

\textbf{\emph{Design.}}
In Section \ref{sec:effectiveness}, we demonstrate that pre-trained models can achieve remarkable performance in vulnerability repair with CVEfixes and Big-Vul datasets (e.g., CodeT5 fixes 44.96\% of vulnerability functions, improving the state-of-the-art technique VRepair by 22.23\%).
However, previous work \cite{chen2022neural,chi2022seqtrans} reports that vulnerability fix datasets remain too small to train a reliable NMT model and generic knowledge learned from related tasks (e.g., bug fixing) can be transferred into the vulnerability repair task.
\revise{Security vulnerabilities can be considered as a subset of software bugs and they are similar in both repair workflows and code patterns.}
To explore the feasibility of transfer learning from \delete{the bug fixing task}\revise{bug fixing} to \delete{the vulnerability repair task}\revise{vulnerability repair} in the pre-trained model scenario, we perform a further investigation to analyze repair performance with and without a bug fixing corpus.

In particular, we design three training scenarios for each pre-trained model.
The \textit{bug fixing scenario} denotes the performance of the model only trained with a corpus of generic bug fixing from Chen et al. \cite{chen2022neural}.
The \textit{vulnerability repair scenario} denotes the performance of the model only trained with our vulnerability fixing corpus.
The \textit{transfer learning scenario} denotes the performance of the model trained with transfer learning, i.e., taking the model trained in \textit{the bug fixing scenario} and fine-tuning it with our vulnerability fix corpus.
In total, we extend our experiment to systematically evaluate the \delete{12}\revise{15} variants of pre-trained model-based vulnerability repair approaches, i.e., 3 training scenarios (bug fixing, vulnerability repair, and transfer learning) × \delete{4}\revise{5} \newdelete{model architectures}\newrevise{pre-trained models} (CodeT5, CodeBERT, GraphCodeBERT\delete{ and UniXcoder}\revise{, UniXcoder and CodeGPT}).

\begin{itemize}
    \item \textbf{M1-M3}: A pre-trained CodeT5 model with bug fixing, vulnerability repair and transfer learning  scenarios.
    \item \textbf{M4-M6}: A pre-trained CodeBERT model with bug fixing, vulnerability repair and transfer learning scenarios.
    \item \textbf{M7-M9}: A pre-trained GraphCodeBERT with bug fixing, vulnerability repair and transfer learning scenarios.
    \item \textbf{M10-M12}: A pre-trained UniXcoder with bug fixing, vulnerability repair and transfer learning scenarios.
    \item \revise{\textbf{M13-M15}: A pre-trained CodeGPT with bug fixing, vulnerability repair and transfer learning scenarios.}
\end{itemize}

\begin{figure}[htbp]
    \centering
    \includegraphics[width=0.95\linewidth]{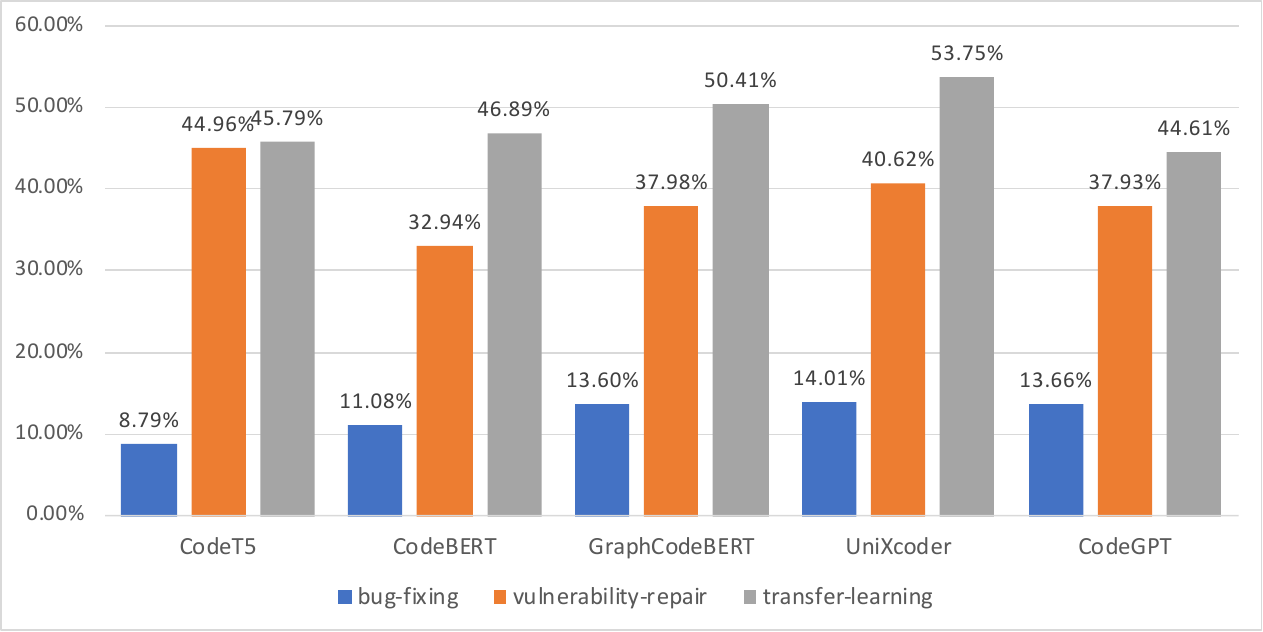}
    \caption{The experimental results of our enhanced pre-trained model-based vulnerability repair approach.}
    \label{fig:transfer}
\end{figure}

\textbf{\emph{Results.}}
The results are presented in Fig. \ref{fig:transfer}.
For each pre-trained model, there exist three bars, indicating the prediction accuracy of the pre-trained model under three training scenarios.
From Fig. \ref{fig:transfer}, we find the transfer learning models achieve the best perfect prediction performance across all pre-trained models with an accuracy of \delete{48.49}\revise{48.29}\% on average.
\delete{Compared with vulnerability repair models, the improvement with transfer learning reaches 0.83\% (45.79\%-44.96\%) for CodeT5,	 13.95\% (46.89\%-32.94\%) for CodeBERT, 12.43\% (50.41\%-37.98\%) for GraphCodeBERT and	13.13\% (53.75\%-40.62\%) for UniXcoder, respectively.}
\revise{Compared with vulnerability repair models, the improvement with transfer learning reaches 0.83\% (45.79\%-44.96\%) for CodeT5,	 13.95\% (46.89\%-32.94\%) for CodeBERT, 12.43\% (50.41\%-37.98\%) for GraphCodeBERT,	13.13\% (53.75\%-40.62\%) for UniXcoder, and 6.68\%~(44.61\%-37.93\%) for CodeGPT, respectively.}
The improvement highlights that the knowledge learned from the bug fixing task is useful and can be transferred to the software vulnerability repair task on pre-trained models.
Notably, compared with the state-of-the-art technique VRepair, transfer learning models can fix more software vulnerability functions.
For example, UniXcoder with transfer learning perfectly predicts vulnerability fixes with an accuracy of 53.75\%, improving VRepair (22.73\%) by 31.02\%.
We also find the bug fixing model can achieve an average accuracy of \delete{11.87}\revise{12.23}\% on the vulnerability repair task.
The results demonstrate that the bug fixing task and the vulnerability repair task share similar code changes.
Thus, future work can be conducted to explore the capabilities of existing bug fixing techniques on repairing software vulnerabilities.

\myfinding{5}{
Our comparison results demonstrate about the approach that:
(1) bug fixing models can repair \delete{11.87}\revise{12.23}\% of software vulnerabilities on average, demonstrating the prospect of existing bug fixing techniques on the vulnerability repair task;
(2) transfer learning from the bug fixing task is able to boost existing vulnerability repair performance, leading to new records on prediction accuracy (e.g., 53.75\% for GraphCodeBERT).
}

\section{Discussion}
\label{sec:dis}

\subsection{Ablation Study of Tags}
\label{sec:tags}

\begin{table*}[htbp]
  \centering
  \caption{\revise{The comparison results of the pre-trained models under different tags.
  The first column list the tag settings and the remaining columns list the corresponding prediction accuracy. 
  We also present the original prediction accuracy with all tags in the last row.}}
    \begin{tabular}{c|ccccc|c}
    \toprule
    \textbf{Type} & \textbf{CodeT5} & \textbf{CodeBERT} & \textbf{GraphCodeBERT} & \textbf{UniXcoder} & \revise{\textbf{CodeGPT}} & \revise{\textbf{Average}} \\
    \midrule
    \textbf{no localization} & 41.03\% & 41.44\% & 37.57\% & 33.41\% & \revise{35.58\%} & \revise{37.81\%} \\
    \textbf{no CWE} & 41.91\% & 37.98\% & 39.21\% & 42.32\% & \revise{30.95\%} & \revise{38.47\%} \\
    \textbf{no tag and cve} & 38.80\% & 37.34\% & 36.11\% & 35.93\% & \revise{32.59\%} & \revise{36.15\%} \\
    \midrule
    \textbf{original} & 44.96\% & 32.94\% & 37.98\% & 40.62\% & \revise{37.93\%} & \revise{38.89\%} \\
    \bottomrule
    \end{tabular}%
  \label{tab:tag}%
\end{table*}%

In this section, we aim to investigate how the special tags \revise{in our datasets} influence the prediction performance of pre-trained models\delete{ for vulnerability repair}.
As discussed in Section \ref{sec:dataset}, we add some special tags (e.g., \textit{$<$StartLoc$>$} and \textit{$<$EndLoc$>$}) to the input and output sequence, which helps the models to localize the vulnerable lines.
These vulnerable lines are identified by off-the-shelf vulnerability analysis tools or security experts.
\delete{For example, a static analyzer (e.g., Infer \cite{infer}) outputs a suspicious line that causes a vulnerability.}
Meanwhile, we also \revise{use} a CWE tag (e.g., CWE-119) at the start of vulnerable functions, indicating what type of CWE category this vulnerability belongs to.
The CWE tag helps the models to learn similar code changes for the vulnerabilities with the same category.
To figure out the contribution of these two types of tags, we respectively remove the localization tag and CWE tag to evaluate the perfect prediction accuracy of pre-trained models.
In total, we extend our experiment to systematically evaluate the \delete{16}\revise{20} variants of pre-trained model-based vulnerability repair approaches, i.e., 4 tag settings (only without localization tag, only without CWE tag, without both types of tags, with both types of tags) × \delete{4}\revise{5} \newdelete{model architectures}\newrevise{pre-trained models} (CodeT5, CodeBERT, GraphCodeBERT\delete{ and UniXcoder}\revise{, UniXcoder and CodeGPT}).

\begin{itemize}
    \item \textbf{M1-M4}: A pre-trained CodeT5 model only without localization tag, only without CWE tag, without both types of tags and with both types of tags.
    \item \textbf{M5-M8}: A pre-trained CodeBERT model only without localization tag, only without CWE tag, without both types of tags and with both types of tags.
    \item \textbf{M9-M12}: A pre-trained GraphCodeBERT only without localization tag, only without CWE tag, without both types of tags and with both types of tags.
    \item \textbf{M13-M16}: A pre-trained UniXcoder model only without localization tag, only without CWE tag, without both types of tags and with both types of tags.
    \item \revise{\textbf{M17-M20}: A pre-trained CodeGPT model only without localization tag, only without CWE tag, without both types of tags and with both types of tags.}
\end{itemize}

The results are presented in Table \ref{tab:tag}.
It can be found that the prediction accuracy of CodeT5 drops by 3.93\% and 3.05\% when the localization tags and CWE tags are removed, respectively.
Meanwhile, CodeT5 achieves 38.80\% without both localization tags and CWE tags, leading to a 6.16\% drop in prediction accuracy.
However, we also find that in some cases, removing the localization or CWE tags may improve the prediction accuracy of CodeBERT, GraphCodeBERT and UniXcoder.
For example, UniXcoder achieves 42.32\% prediction accuracy without CWE tags, which is 1.70\% higher than the model with CWE tags.
Overall, although such tags provide benefits for CodeT5, there exist adverse effects for other models in vulnerability repair.
These results demonstrate that such tags may not provide sustainable benefits for all studied pre-trained models, which is contrary to the conclusion drawn from a vanilla transformer \cite{chen2022neural}.
\revise{
After our careful analysis, we think the possible reason comes from the different model architectures.
When decoder-only models are used to repair security vulnerabilities, a new decoder is fine-tuned from scratch and does not contain any pre-training knowledge, while CodeT5 contains both pre-trained encoder and decoder models to support vulnerability repair naturally. Thus, it is difficult for encoder-only models to understand some additional tag information with a new decoder.
}
Future researchers should explore the actual usage of tags to improve the prediction accuracy of pre-trained vulnerability repair models.

\mydiscussion{
The pre-trained models have diverse repair performance under the tags, e.g., there exist a 6.16\% decrease for CodeT5 whereas a 4.40\% increase for CodeBERT in prediction accuracy when removing all tags.
}

\subsection{Code Representation}
\label{sec:repres}

In this section, we aim to investigate how different code representation influence the prediction performance of pre-trained models.
As discussed in Section \ref{sec:dataset}, we adopt a tag-based code change description for vulnerability repair, which allows for sequences shorter than a whole function to be generated.
According to previous repair work \cite{tufano2019empirical,yuan2022circle}, we consider two potential code representation additionally\delete{: diff-based and prompt-based}.

\textit{(1) The diff-based representation.}
Most existing NMT-based repair work \cite{tufano2019empirical} simply takes a whole vulnerable function as input and generates a patched function\delete{, and the performance is found to decrease as the function length increases}.
However, we found that the vulnerable functions in our dataset have over 900 tokens on average, which is significantly larger than the dataset in \cite{tufano2019empirical} (less than 100 tokens).
Thus, our vulnerability dataset cannot be directly applied to such code representation as the transformer limits the maximum length of the input sequence to $512$ tokens.
To reduce the function length, we perform a diff operation between a vulnerable and correct function.
We then extract the removed and added lines as the vulnerable and patched lines, marked with ``{\color{green}{+}}'' and ‘{\color{red}{-}}’, respectively.
\textit{(2)The prompt-based representation.}
\delete{Meanwhile, }Raffel et al. \cite{raffel2019t5} propose a text-in-text-out input format, which concatenates different input components with some prefixed prompt.
This prompt mechanism is proved to be useful for fine-tuning pre-trained models in downstream tasks. 
To evaluate how a prompt-based \delete{code }representation performs for pre-trained models in vulnerability repair, we adopt a manually designed prompt template to convert a vulnerable function into a unified fill-in-the-blank format proposed in previous repair work \cite{yuan2022circle}.
In total, we extend our experiment to systematically evaluate the \delete{12}\revise{15} variants of pre-trained model-based vulnerability repair approaches, i.e., 3 representation settings (tag-based, diff-based and prompt-based) × \delete{4}\revise{5} \newdelete{model architectures}\newrevise{pre-trained models} (CodeT5, CodeBERT, GraphCodeBERT\delete{ and UniXcoder}\revise{, UniXcoder and CodeGPT}).

\begin{itemize}
    \item \textbf{M1-M3}: A pre-trained CodeT5 model with tag-based, diff-based and prompt-based code representation.
    \item \textbf{M4-M6}: A pre-trained CodeBERT model with tag-based, diff-based and prompt-based representation.
    \item \textbf{M7-M9}: A pre-trained GraphCodeBERT with tag-based, diff-based and prompt-based representation.
    \item \textbf{M10-M12}: A pre-trained UniXcoder model only with tag-based, diff-based and prompt-based representation.
    \item \revise{\textbf{M13-M15}: A pre-trained CodeGPT model only with tag-based, diff-based and prompt-based representation.}
\end{itemize}

\begin{figure}[t]
    \centering
    \includegraphics[width=0.95\linewidth]{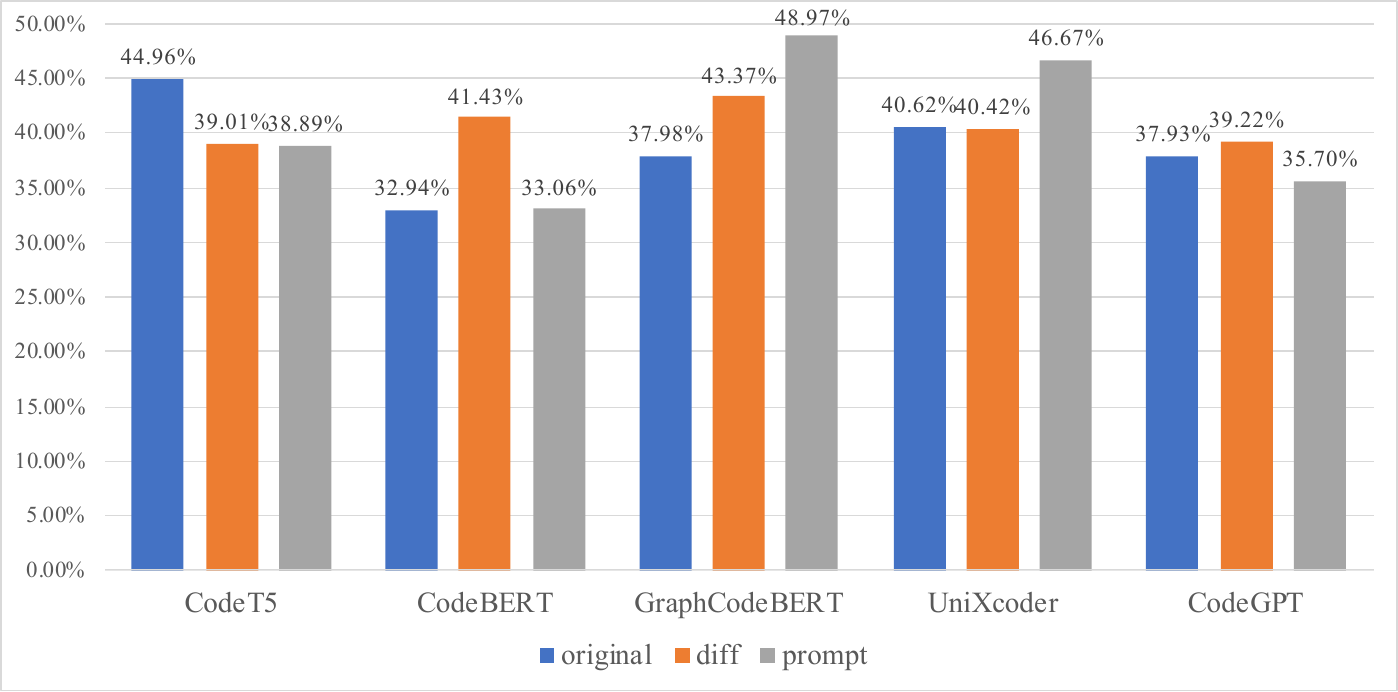}
    \caption{\revise{The experimental results of pre-trained models with different representation for vulnerability repair.}}
    \label{fig:prompt}
\end{figure}

The results are presented in Fig. \ref{fig:prompt}.
When considering diff-based representation, we find it achieves \delete{41.69}\revise{40.69}\% prediction accuracy on average, \delete{2.56}\revise{1.80}\% higher than tag-based representation.
In particular, it can be found that diff-based representation outperforms tag-based representation for CodeBERT and GraphCodeBERT by 8.49\%~(41.43\%-32.94\%) and 5.39\%~(43.37\%-37.98\%), while tag-based representation achieves better prediction accuracy for CodeT5.
When \delete{comparing}\revise{considering} prompt-based representation, pre-trained models predict correct fixes for \delete{42.58}\revise{40.66}\% of vulnerabilities, and the improvement reaches \delete{3.45}\revise{1.77}\%.
In particular, GraphCodeBERT successfully repairs 48.97\% of vulnerability functions with prompt-based representation, \revise{10.99\%~(48.97\%-37.98\%)} higher than tag-based representation.
These results demonstrate that prompt representation can perform well for encoder-only pre-trained models in vulnerability repair.
Thus, future researchers should further explore more prompts that can utilize generic pre-training knowledge well to improve the accuracy of \delete{the NMT-based }vulnerability repair models.
\revise{Besides, considering there exist a number of studies to represent source code in deep-learning code-related tasks ~\cite{zhang2019novel,ben2018neural,ye2020deep,li2019improving}, it is promising to conduct a systematic study to investigate the impact of  various code representation in pre-trained model-based vulnerability repair.}

\mydiscussion{
When applying pre-trained models to vulnerability repair, the diff-based representation achieves comparable prediction accuracy against the prompt-based one, while both of them outperform the original tag-based representation, with \revise{1.80\% and 1.77\%} improvement on average in terms of prediction accuracy.
}

\subsection{\revise{Preliminary Study of ChatGPT}}
\label{sec:gpt}
\revise{
Recently, ChatGPT~\cite{chatgpt}, a very powerful pre-trained model developed by OpenAI based on the GPT architecture, has attracted increasing attention and shown an impressive ability to address various tasks. 
Different from most pre-trained models (e.g., CodeT5), ChatGPT is trained with reinforcement learning from human feedback (RLHF) and a significantly larger model scale.
However, it still remains unclear how effective ChatGPT is in repairing security vulnerabilities.
We exclude ChatGPT in our experiment (discussed in Section~\ref{sec:model}) as ChatGPT is close-sourced\delete{ and can only be accessed by API}.
In this section, we preliminarily perform the first study to investigate how ChatGPT performs in repairing security vulnerabilities.
We interact with ChatGPT through natural language conversations, i.e., sending requests (i.e., the designed prompt\footnote{Prompt in our study: You are an automated vulnerability repair tool. The following C function contains some vulnerable lines. Each vulnerable line is identified by the tags $\langle \texttt{ModStart}\rangle$ and $\langle \texttt{ModEnd}\rangle$. Please provide a correct function: [the vulnerable function code block].} with the vulnerable code) to ChatGPT and receiving responses (i.e., correct code) from ChatGPT in a zero-shot manner.}

\revise{
In particular, we access ChatGPT with its API of \textit{gpt-3.5-turbo-0301}, which is the latest
version available. 
We randomly select 100 samples from our dataset to perform this preliminary study due to the limitation of free code tokens in ChatGPT.
For each sample, we run ten independent requests by starting a new conversation every time on the API due to the non-deterministic nature of ChatGPT.
As a result, we find only nine out of 100 samples are correctly fixed by ChatGPT, underperforming pre-trained models investigated in Section~\ref{sec:re&an}.
After our careful analysis, the possible reason lies in the design and application scenario of ChatGPT.
ChatGPT is designed to address lots of tasks with all possible natural languages and programming languages in the wild.
On the contrary, existing pre-trained models are particularly fine-tuned with the vulnerability dataset, and are able to capture the transformation rules to repair security vulnerabilities.
We then manually inspect the generated patches from ChatGPT and find there are five patches semantically equivalent to the ground truth.
We think the possible reason is that ChatGPT is able to understand the semantics of the vulnerable code, and attempts to generate correct code.
Overall, our findings indicate that ChatGPT contains extensive knowledge of programming language while failing to understand the intention of vulnerability.
As a result, it is challenging to directly employ ChatGPT to repair security vulnerabilities in a zero-shot manner.
Future researchers may design specific ChatGPT-based vulnerability repair techniques, e.g., providing vulnerability description information in the form of conversational interactions.
}

\mydiscussion{
Despite encompassing vast knowledge of programming languages, ChatGPT fails to demonstrate promising repair performance without fine-tuning.
}

\section{Implication and Guideline}
\label{sec:implication}

\delete{Based on the observations in our experiment, we}\revise{We} summarize the following \delete{essential }practical guidelines for future \revise{pre-trained model-based} vulnerability repair studies.

\textbf{The advance of pre-trained models.}
\delete{Our~}\revise{In Section~\ref{sec:effectiveness}, our} results show that the studied \delete{four}\revise{five} pre-trained models perform even better than the state-of-the-art technique VRepair.
Besides, transferring learning from bug fixing can further enhance the repair performance \revise{(mentioned in Section~\ref{sec:rq5})}.
Such observation motivates future researchers to investigate more advanced vulnerability repair  techniques by employing different pre-trained models.
For example, it is interesting to propose domain-specific pre-trained models by designing vulnerability repair-related pre-training tasks.
Meanwhile, thorough evaluations are recommended to explore how different features, such as CWE types and CVE severity scores, influence the performance of pre-trained models in vulnerability repair.

\revise{
\textbf{The improvement of fine-tuning.}
In Section~\ref{sec:rq3}, fine-tuned pre-trained models with more vulnerability samples always lead to better repair performance.
We can find there is an approximately linear relationship between repair accuracy and fine-tuning data size in Fig~\ref{fig:fine-tuning}.
The accuracy growth trend even has not slowed down with all data in our experiment, suggesting the possibility of better performance with more fine-tuning datasets.
Besides, the fine-tuning method is straightforward and simple due to the empirical nature of our work.
In the future, it is promising to incorporate some repair-specific designs to fine-tune pre-trained models, such as code edits~\cite{zhu2021syntax} and dynamic execution information~\cite{chen2022neural}.
}

\textbf{The benefits of bug fixing.}
\delete{Our~}\revise{In Section~\ref{sec:rq5}, our} study demonstrates that NMT-based models only trained on a limited bug fixing corpus can already fix notable vulnerabilities.
Besides, NMT-based repair models with a transfer learning paradigm\delete{, which is first trained on a bug fixing corpus and then tuned on a vulnerability-fixing dataset,} significantly improve the prediction accuracy\delete{ (e.g., the improvement reaches \delete{10.00}\revise{9.40}\% on average in Fig. \ref{fig:transfer})}. 
These results indicate that bug fixing and vulnerability repair both aiming to fix errors in the source code have a high degree of similarity and the knowledge learned from bug fixing can be well transferred to vulnerability repair.
However, these two tasks are developing in their respective fields so far and little work has explored their potential relationship.
\revise{Different from common software bugs,  security vulnerabilities are a subset of bugs and are more detrimental and necessitate immediate remedies.}
Such observation motivates future researchers to investigate whether the two tasks can benefit each other, e.g., how to migrate advanced bug fixing techniques to automated vulnerability repair.

\section{Threats to validity}
\label{sec:threats}

\delete{To facilitate the replication and verification of our experiments, we have made the relevant materials (including source code, trained models, and vulnerability dataset) publicly available~\cite{myurl}.}
\delete{Despite that, our empirical study still faces some threats to validity, summarized as follows.}

The first threat to validity lies in the vulnerability benchmarks.
In our experiment, we focus on the CVEfixes and  Big-Vul corpora with 8,482 security vulnerabilities from 180+ different CWE types in total according to previous studies \cite{chen2022neural, fu2022vulrepair}.
It is unclear the degree to which our conclusions can be generalized to other corpora and CWE types.
Considering that the two corpora are collected  in two independent papers \cite{bhandari2021cvefixes, fan2020ac}, we have high confidence in the generalizability of our conclusions. 
Besides, in future work, we will explore the actual advantage of pre-trained models with more datasets for vulnerability repair.

The second threat to validity is that our findings may not generalize to other pre-trained models.
We select CodeT5, CodeBERT, GraphCodeBERT\delete{ and UniXcoder}\revise{, UniXcoder and CodeGPT} in our experiment due to their powerful performance in recent code-related works \cite{guo2022unixcoder, wang2021codet5}.
However, it is unclear whether the conclusions in our experiment (discussed in Section \ref{sec:re&an}) can be maintained when using other pre-trained models.
We have mitigated the potential threat by using encoder-decoder, encoder-only and \delete{unified}\revise{decoder-only} architectures to demonstrate the performance of diverse pre-trained models.

\revise{
The third threat to validity comes from the selected baseline in RQ1.
In Section~\ref{sec:effectiveness}, we select VRepair as the compared technique to evaluate the effectiveness of pre-trained models following recent work~\cite{fu2022vulrepair}.
We fail to consider other techniques due to several reasons, detailed in Section~\ref{sec:related}.
However, considering that VRepair is the most recent technique and our work mainly focuses on empirical evaluations, the improvement of the selected models over VRepair is enough to demonstrate the promising future of boosting vulnerability repair on top of pre-trained models.
}

The last threat to validity is the evaluation metric.
In our work, we evaluate the repair performance with \delete{a perfect match (i.e., whether the generated sequences is the same as the correct sequences)}\revise{prediction accuracy}.
Previous work \cite{qi2015analysis, yi2018correlation} demonstrates that code repair approaches should be evaluated on both plausible patches (i.e., \delete{a candidate patch }passing the available test suite) and correct patches (i.e., \delete{a plausible patch} semantically equivalent to the developer patch by manual inspection).
However, these two metrics require the test suite to identify whether the repair candidates can successfully pass the test suite or not. 
Unfortunately, we fail to utilize the two metrics as there exists no test suite in the CVEFixes and Big-Vul datasets.
Thus, following existing work \cite{chen2022neural,fu2022vulrepair,chi2022seqtrans}, we utilize perfect prediction to evaluate the vulnerability repair task.
Future researchers could create new vulnerability fixing datasets with corrected test suites.

\section{Related Work}
\label{sec:related}

In this section, we discuss the scope of our work with respect to NMT-based vulnerability repair and pre-trained models.

\subsection{NMT-based Automated Vulnerability Repair}
With the advancement of NMT techniques, researchers propose NMT-based automated vulnerability repair approaches~\cite{zhang2023survey}.
For example, Chen et al. \cite{chen2022neural} propose VRepair based on a vanilla transformer model, which is first pre-trained on a bug fixing corpus and then fine-tuned for the vulnerability repair task.
Similarly, Chi et al. \cite{chi2022seqtrans} propose a transformer-based technique SeqTrans to repair Java securities, which adopts the same pre-training and fine-tuning components as VRepair.
It's worth noting that SeqTrans leverages code abstraction to reduce the vocabulary size and make the NMT model focus on learning common repair patterns from different code changes.
Fu et al. \cite{fu2022vulrepair} propose a T5-based automated vulnerability repair approach VulRepair that leverages the pre-training and BPE components.
\revise{Pearce et al.~\cite{pearce2022examining} employ black-box pre-trained models to produce security fixes in a zero-shot scenario, while our work focuses on open-source pre-trained models with additional fine-tuning, leading to two orthogonal dimensions.}

Unlike existing NMT-based vulnerability repair work, we perform the first systematic experiment to investigate the performance of pre-trained models on the vulnerability repair task.
The comprehensive evaluations demonstrate the substantial improvement of pre-trained models compared with VRepair, highlighting the significant advancement of using pre-trained models to repair security vulnerabilities.

\subsection{Pre-trained Models}
\delete{
In this section, we first introduce the existing studies about pre-trained models in NLP.
We then discuss the application of pre-trained models in some typical code-related tasks and other domains about programming languages.}

\delete{\textit{Pre-trained models in NLP.}}
Recent work has demonstrated substantial gains on various NLP tasks by pre-training on a large corpus of text followed by fine-tuning on a specific task.
For example, Devlin et al. \cite{devlin2018bert} propose a novel language representation model BERT to pre-train deep bidirectional representation from the unlabeled text by jointly conditioning on both left and right contexts in all layers.
To explore the landscape of transfer learning\delete{ techniques for NLP}, Raffel et al. \cite{raffel2019t5} propose a text-to-text transfer transformer T5 by introducing a unified framework that converts all text-based language problems into a text-to-text format.
\looseness=-1

\delete{\textit{Pre-trained models in code-related tasks.}}
Inspired by the success of pre-trained models in NLP, many researchers have proposed applying pre-trained models to code-related tasks.
Instead of designing new network architectures, researchers usually adopt existing architectures in NLP\delete{ (e.g., a vanilla transformer \cite{vaswani2017attention})} and design some code-aware pre-training tasks (e.g., code-AST prediction \cite{wang2021codet5}) to learn structure representation of the source code.
Then pre-trained models are further fine-tuned to some diversified code-related tasks such as code-code (e.g., code translation), text-code (e.g., code generation), and code-text (e.g., code summarization) scenarios \cite{lu2021codexglue}.
We have already introduced some state-of-the-art code-related pre-trained models in Section \ref{sec:model}.
\looseness=-1

\delete{\textit{Applications of pre-trained models.}}
In addition to the above-mentioned typical tasks (e.g., \delete{code mutant injection and }code summarization in \cite{mastropaolo2021t5learning}), researchers apply pre-trained models to other code-related domains.
Tufano et al. \cite{tufano2022using} propose the usage of a pre-trained T5 model relying on a SentencePiece tokenizer to boost the code review task.
Yuan et al. \cite{yuan2022circle} propose CIRCLE, a T5-based program repair framework equipped with continual learning ability across multiple languages.
Zhang et al. \cite{zhang2023gamma} propose GAMMA, a template-based program repair approach with mask prediction on top of the large language models.
\revise{
Recently, Jiang et al.~\cite{jiang2023impact} explore the performance of pre-trained models with fine-tuning for the program repair domain.
Different from that work, we conduct a comprehensive study to study pre-trained models from a systematic perspective for repairing security vulnerability, i.e., the repair pipeline including data pre-processing, model training, and repair inference.
}

Although there exist some code-related tasks (e.g., code review) that already benefit from pre-trained models, in this work, we perform the first extensive study to investigate how pre-trained models are applied in vulnerability repair.

\section{Conclusion}
\label{sec:conclusion}

In this paper, we perform the first exploration to investigate the feasibility of pre-trained models on automated vulnerability repair.
Specifically, we conduct a large-scale study to analyze the  effectiveness and limitations of vulnerability repair approaches supported by pre-trained models, involving two vulnerability datasets and more than 100 trained models.
The results demonstrate that pre-trained models can consistently outperform the state-of-the-art technique VRepair by \delete{16.40}\revise{16.16}\% on average, highlighting the substantial advancement of pre-trained models in vulnerability repair.
We further analyze the influence of different aspects in the repair workflow, such as dataset pre-processing and model training.
Surprisingly, we also demonstrate that a simplistic pre-trained model-based repair approach, which uses the transfer learning from bug fixing, can further boost repair performance.
Lastly, our findings reveal various practical guidelines for future pre-trained model-based vulnerability repair.
Overall, our work demonstrates the promising future of using pre-trained models to generate real-world correct vulnerability patches and help under-resourced security analysts to fix security vulnerabilities in practice.

\ifCLASSOPTIONcompsoc
  \section*{Acknowledgments}
\else
  \section*{Acknowledgment}
\fi

This work is supported partially by the National Natural Science Foundation of China (61932012, 62141215, 62372228) and Science, Technology and Innovation Commission of Shenzhen Municipality (CJGJZD20200617103001003).
\ifCLASSOPTIONcaptionsoff
  \newpage
\fi

\bibliographystyle{IEEEtran}
\bibliography{reference}

\vspace{-5 mm}
\begin{IEEEbiographynophoto}
{Quanjun Zhang}
is currently working toward the Ph.D. degree in Software Institute at Nanjing University, Nanjing, China. 
His current research interests include intelligent software testing and program repair.

\vspace{10pt}
\noindent 
\textbf{Chunrong Fang}
received the B.E. and Ph.D. degrees in software engineering from Software Institute, Nanjing University, Jiangsu, China. 
He is currently an assistant professor with the Software Institute of Nanjing University.
His research interests lie in intelligent software engineering, e.g. BigCode and AITesting.

\vspace{10pt}
\noindent 
\textbf{Bowen Yu}
is currently working toward the B.E. degree in Software Institute at Nanjing University, Nanjing, China. 
His current research interests include automated program repair and large language models.

\vspace{10pt}
\noindent 
\textbf{Weisong Sun}
is the Ph.D. Candidate of Software Institute, Nanjing University, China. His research interests include Intelligent Software Engineering and Trustworthy Artificial Intelligent.

\vspace{10pt}
\noindent 
\textbf{Tongke Zhang}
is currently working toward the B.E. degree in Software Institute at Nanjing University, Nanjing, China. 
His current research interests include automated program repair and large language models.

\vspace{10pt}
\noindent 
\textbf{Zhenyu Chen}
is currently a full professor with Software Institute of Nanjing University. 
He is an associate Editor of IEEE Transactions on Reliability. He is also the Contest Co-Chair at QRS 2018, ICST 2019, and ISSTA 2019. 
He is the Industrial Track Co-Chair of SANER 2019. 
His research interests include collective intelligence, deep learning testing and optimization, big data quality, and mobile application testing.
\end{IEEEbiographynophoto}

\vfill

\end{document}